\documentclass[prd,twocolumn,amsmath,amssymb,floatfix,superscriptaddress,nofootinbib]{revtex4}

\usepackage{amsmath}
\usepackage{epsfig}
\usepackage{color}
\usepackage{natbib}
\usepackage{graphicx}

\newcommand{\ModeCode}{{\sc Mode\-Code}}
\newcommand{\MultiNest}{{\sc MultiNest}}

\newcommand{\CosmoMC}{{\sc CosmoMC}}

\begin{document}

\title{Bayesian Analysis of Inflation III: Slow Roll Reconstruction Using Model Selection}

\author{Jorge Nore\~na}\email{jorge.norena@icc.ub.edu}\affiliation{ICC, University of Barcelona (IEEC-UB), Marti i Franques 1, Barcelona 08028, Spain \\ \mbox{}}

\author{Christian Wagner}\email{cwagner@icc.ub.edu}\affiliation{ICC, University of Barcelona (IEEC-UB), Marti i Franques 1, Barcelona 08028, Spain \\ \mbox{}}

\author{Licia Verde}\email{liciaverde@icc.ub.edu}\affiliation{ICC, University of Barcelona (IEEC-UB), Marti i Franques 1, Barcelona 08028, Spain \\ \mbox{}}\affiliation{ICREA, Instituci\'o Catalana de Recerca i Estudis Avan\c{c}ats \\ \mbox{}}

\author{Hiranya V. Peiris}\email{h.peiris@ucl.ac.uk}\affiliation{Department of Physics and Astronomy, University College London, London WC1E 6BT, U.K. \\ \mbox{}}

\author{Richard Easther}\email{r.easther@auckland.ac.nz}\affiliation{Department of Physics, Yale University, New Haven, CT 06520, U.S.A.}
\affiliation{Department of Physics, University of Auckland, Private Bag 92019, Auckland,
New Zealand  \\ \mbox{}}

\date{\today}

\begin{abstract}
\baselineskip 11pt

We implement Slow Roll Reconstruction -- an optimal solution to the inverse problem for inflationary cosmology -- within \ModeCode, a publicly available solver for the inflationary dynamics.  We obtain up-to-date constraints on the reconstructed inflationary potential, derived from  the WMAP 7-year dataset and  South Pole Telescope observations, combined with large scale structure data derived  from SDSS Data Release 7.   Using \ModeCode\ in conjunction with the \MultiNest\  sampler, we compute Bayesian evidence for the reconstructed potential at each order in the truncated slow roll hierarchy.    We find that the data are well-described by the first two slow roll parameters, $\epsilon$ and $\eta$, and  that there is no need to include a nontrivial $\xi$ parameter.
\end{abstract}

\maketitle

\section{Introduction}

Inflation \cite{Guth:1980zm,Linde:1981mu} postulates that the very early universe underwent a period of accelerated expansion. It is typically  described via the dynamics of the inflaton, a scalar effective  degree of freedom coupled to gravity. Quantum fluctuations of the inflaton and other light fields  constitute the most widely studied mechanism for producing  primordial density perturbations (for reviews see Refs.~\cite{Baumann:2009ds,Lyth:1998xn}).  These perturbations induce temperature fluctuations in the cosmic microwave background radiation (CMB) and, as a consequence of their gravitational growth,  lead to the formation of the large scale structure of the universe (LSS). 

A key corollary of the inflationary hypothesis  is that precise measurements of the CMB and the distribution of galaxies in the sky constrain the physical mechanism that drives the accelerated expansion, opening a remarkable window into the physics of the very early universe.   In recent years, high-quality observations of the CMB have been provided by the WMAP satellite~\cite{Komatsu:2010fb},  ACBAR~\cite{acbar},  the South Pole Telescope (SPT)~\cite{Keisler:2011aw} and the Atacama Cosmology Telescope (ACT) \cite{act}, while the Sloan Digital Sky Survey (SDSS) \cite{Abazajian:2008wr} has probed the density perturbations at low redshifts. Further,  data from the  Planck  satellite \cite{:2006uk} will soon be available to the cosmological community, and several major LSS surveys are now underway or are being planned.

Inflation can be realized in a vast number of ways, and these scenarios are typically distinguishable via their predictions for the statistical properties of the primordial perturbations. Some models are purely {\em ad hoc\/}, while others satisfy naturalness constraints, or can be derived from candidate  theories of fundamental physics. Given the plethora of models that exist, it is fruitful to consider the corresponding inverse problem: given cosmological data, can we deduce the mechanism underlying inflation?  This problem is often referred to as {\em reconstruction\/}  \cite{Turner:1993su,Copeland:1993jj,Copeland:1993ie,Liddle:1994cr,Lidsey:1995np}.   Ideally, reconstruction would  recover the effective action of the ``inflationary sector'' of high energy physics, but carrying out this program for a fully general scenario is not likely to be feasible, even with  data from the next generation of astrophysical experiments.  Consequently,  present-day implementations of reconstruction incorporate a (sometimes implicit) prior restricting attention to a specific category of models -- usually a single inflaton with a canonical kinetic term, minimally coupled to Einstein gravity.   

This paper builds on the Slow Roll Reconstruction  algorithm, proposed  by Easther and Peiris \cite{Peiris:2006sj,Peiris:2006ug,Easther:2006tv,Peiris:2008be,Adshead:2008vn}.   Slow Roll Reconstruction does not require the slow roll {\em approximation}, but rather relies on the truncated Hamilton-Jacobi slow roll {\em expansion}  \cite{Lidsey:1995np,Muslimov:1990be,Salopek:1990jq,Salopek:1990re,Lidsey:1991zp,Liddle:2003py}. This is a phenomenological description of inflation, obtained by expanding the Hubble parameter as a power series, with the inflaton field as the independent variable or ``clock''.  We  focus  on scales which are directly probed by observations,  making minimal assumptions  regarding the reheating mechanism and the expansion history of the universe during the ``primordial dark age''  \cite{Boyle:2005se}.    

Given the long history of the inflationary inverse problem, Slow Roll Reconstruction has a number of antecedents.  In particular, other approaches to the inverse problem  based on the Hamilton-Jacobi equations include Ref. \cite{Hoffman:2000ue} and ``Monte Carlo Reconstruction'' \cite{Kinney:2002qn,Easther:2002rw}, which was further developed in Refs~\cite{Kinney:2006qm,Powell:2007gu,Powell:2008bi}.   Separately, Ref.~\cite{Malquarti:2003ia}  constrained the slow roll
parameters by requiring that the duration of inflation was sufficient to solve the classic cosmological problems, while Cline and Hoi reconstructed inflationary models that allow for a significant running of the scalar spectral index within the Hamilton-Jacobi formalism~\cite{Cline:2006db}.  Leach and collaborators  constrained inflation by writing the spectral indices in terms of the slow roll parameters  \cite{Leach:2002ar,Leach:2002dw, Leach:2003us}.   A similar scheme to Slow Roll Reconstruction was discussed in Refs.~\cite{Lesgourgues:2007aa,Hamann:2008pb}. Likewise, Ref.~\cite{Finelli:2009bs} used the WMAP5 dataset and  SDSS data release 7 LRG data to constrain the Hubble slow roll parameters, while Ref. \cite{Kawasaki:2009yn}  obtained constraints on the power spectra of curvature and tensor perturbations using priors based on single field slow roll inflationary models.

We implement Slow Roll Reconstruction within \ModeCode\footnote{http://zuserver2.star.ucl.ac.uk/$\sim$hiranya/ModeCode}, a publicly available solver for the inflationary background  and perturbations \cite{Mortonson:2010er,Easther:2011yq}.  As shown by Liddle \cite{Liddle:2003py} the background dynamics corresponding to the truncated slow roll hierarchy can be solved analytically, yielding the corresponding inflationary potential $V(\phi)$.  Adding this potential to the menu of models  supported within   \ModeCode\ yields a simple and robust implementation of  Slow Roll Reconstruction. Further, \ModeCode\  performs a full numerical computation of the inflationary perturbation spectrum,   making no use of the slow roll approximation when computing the power spectrum. We then estimate the slow roll parameters using the WMAP7 data, and the recent SPT and SDSS DR7 data releases.  

The principal advantage of implementing   Slow Roll Reconstruction within \ModeCode\  is that, by using the nested sampler  \MultiNest\ \cite{Feroz:2007kg,Feroz:2008xx} with \CosmoMC\ \cite{Lewis:2002ah}, we compute Bayesian evidence at each order in the truncated slow roll hierarchy.  This information  determines the number of slow roll parameters that are required to account for the data, and thus the optimal order at which to truncate the hierarchy.  In addition, we carefully construct the priors for the slow roll parameters to ensure that the computed evidence values are realistic \cite{Easther:2011yq}.   Finally, we compute two heuristic model selection statistics -- the Profile Likelihood ratio and the Akaike Information Criterion -- for the truncated slow roll hierarchy, and compare these to the Bayesian inferences.

This  is the third in a sequence of papers on Bayesian analysis of inflation, and optimal approaches to constraining inflationary models with astrophysical data. The first \cite{Mortonson:2010er} focussed on estimating the free parameters in specific inflationary models, while the second \cite{Easther:2011yq} addressed the model selection problem in inflation and the computation of Bayesian evidence.

The paper is organized as follows: In Section \ref{sec:SRR} we describe  Slow Roll Reconstruction and summarize its strengths.  Section \ref{sec:method} contains a detailed  description of our analysis, and  summarizes the  data used to generate the constraints. We present our results in Section \ref{sec:results}, and summarize the implications of our findings in Section \ref{sec:conclusions}. 

\section{Slow Roll reconstruction}\label{sec:SRR}

During inflation the background metric is well-described by the flat Friedmann Robertson Walker metric
\begin{equation}
\mathrm{d}s^2 = -\mathrm{d}t^2 + a(t)^2\mathrm{d}x^2\,,
\end{equation}
where $a(t)$ is the scale factor and, as usual,  the Hubble parameter $H \equiv \dot a / a$ where dots denote derivatives with respect to time, $t$. Inflation is, by definition, a period of accelerated expansion during which  $\ddot a > 0$.  

Slow Roll Reconstruction \cite{Peiris:2006sj,Peiris:2006ug,Easther:2006tv,Peiris:2008be,Adshead:2008vn} is based on the inflationary flow hierarchy.  This is obtained by rewriting the second Friedmann equation,  $\dot{H} = - \dot\phi^2/(2 M_\mathrm{pl}^2)\,$, as
\begin{equation}
\dot{\phi} = -2M_\mathrm{pl}^2H'(\phi)\,,
\end{equation}
where  $M_\mathrm{pl}$ is the reduced Planck mass and primes denote derivatives with respect to $\phi$.  We  define the slow roll parameters, ${}^{(n)}\!\lambda\,$:
\begin{equation}
^{(0)}\lambda= 2 M_{\rm Pl}^2\left[\frac{H'(\phi)}{H(\phi)}\right]^2 \equiv \epsilon
\end{equation}
for $n = 0$, and 
\begin{equation}
{}^{(n)}\!\lambda \equiv (2 M_\mathrm{pl}^2)^n \frac{(H')^{n-1}}{H^n}\frac{\mathrm{d}^{n+1}H}{\mathrm{d}\phi^{n+1}}
\end{equation}
for $n\ge 1$.
These are related to the Hubble slow roll parameters by noting that $\eta \equiv {}^{(1)}\!\lambda$ and $\xi \equiv {}^{(2)}\!\lambda$ \cite{Liddle:1994dx}. We  obtain the Hamilton-Jacobi equations  \cite{Muslimov:1990be,Salopek:1990jq,Salopek:1990re,Lidsey:1991zp} by differentiating ${}^{(n)}\!\lambda$ with respect to $\phi$, yielding an infinite hierarchy of differential equations where ${}^{(n+1)}\!\lambda$ is determined by the two previous terms in the expansion. If ${}^{(n)}\!\lambda \equiv0$ for all $n>N$ at a specific value of $\phi$, the hierarchy truncates, and these terms must be zero for all values of $\phi$. As pointed out by Liddle \cite{Liddle:2003py}, this ensures that the higher-order derivatives of $H(\phi)$  vanish for all $\phi$,  which is equivalent to requiring that $H(\phi)$ is a polynomial of finite order,
\begin{equation}
H(\phi) = H_\ast \big(1 + A_1 \phi + A_2 \phi^2 + \dots + A_N \phi^N\big)\,,
\label{eq:SRRdef}
\end{equation}
where a subscript $\ast$ denotes a quantity evaluated at $\phi = 0$.  Multiplying $H$ by a constant leaves the  ${}^{(n)}\!\lambda$ unchanged;   as we will see below, $H_\ast$ corresponds to the overall energy scale of the inflationary era. The coefficients $A_1$, $\dots$, $A_N$ in equation (\ref{eq:SRRdef}) are then related to the Hubble slow roll parameters by  
\begin{eqnarray}
A_1 &=& \bigg(\frac{\epsilon_\ast}{2 M_\mathrm{pl}^2}\bigg)^{1/2}\,, \label{eq:A1}\\
A_n &=& \bigg(\frac{1}{2M_\mathrm{pl}^2}\bigg)^{n - 1} \frac{A_1^{n-1}}{n!}{}^{(n-1)}\!\lambda_\ast\,.\label{eq:An}
\end{eqnarray}
The system is not modified  by the shift $\phi\rightarrow \phi+\phi_0$, but this rescaling does implicitly redefine the $A_n$. We remove this ambiguity without loss of generality by assuming that the slow roll parameters are measured  at the instant at which $\phi=0$.   Further, the slow roll hierarchy simplifies when $\epsilon \ll 1\,$, leading to a distinct class of solutions, and we  test both ``High-$\epsilon$ N-parameter"  and ``Low-$\epsilon$ N-parameter" models \cite{Adshead:2008vn} in what follows.

Given an expression for $H(\phi)$ we can always obtain the corresponding potential $V(\phi)$ \cite{Muslimov:1990be,Salopek:1990jq,Salopek:1990re,Lidsey:1991zp,Liddle:2003py} by recalling that   $H^2 = ( \dot\phi^2 /2 + V(\phi))/3M_\mathrm{pl}^2$ and replacing $\dot{\phi}$ with $H'$, 
\begin{multline}
V(\phi) = M_\mathrm{pl}^2H_\ast^2\big[3(1+A_1\phi+\dots+A_N\phi^N)^2 \\- 2(A_1+\dots+NA_N\phi^{N-1})^2\big]\,.\label{eq:V}
\end{multline}
 This  expression is a function of the $A_n$ and thus the slow roll parameters, so  Slow Roll Reconstruction can be implemented by adding equation~(\ref{eq:V}) to the ``menu'' of inflationary models  supported by \ModeCode\ \cite{Mortonson:2010er,Easther:2011yq}.  

The Fourier components of the scalar and tensor modes $u_k$ and $v_k$  are  obtained by solving
\begin{align}
u_k{}_{,\tau\tau} + \bigg(k^2 + \frac{z_{,\tau\tau}}{z}\bigg) u_k &= 0\,, \\
v_k{}_{,\tau\tau} + \bigg(k^2 - \frac{a{}_{,\tau\tau}}{a}\bigg) v_k &= 0\,,
\end{align}
where $z\equiv \dot\phi/H$ and the subscript $\tau$ denotes a derivative with respect to conformal time $\mathrm{d}\tau \equiv \mathrm{d}t/a$. Initial conditions for these modes are set by the Bunch-Davies vacuum when the mode $k$ is well within the horizon, and the amplitude of the power spectrum, defined by
\begin{equation}
\langle\mathcal{R}_\mathbf{k} \mathcal{R}_\mathbf{k'}\rangle = \frac{2\pi^2}{k^3}\Delta^2_\mathcal{R}(k) (2\pi)^3 \delta^{(3)}(\mathbf{k} - \mathbf{k'})\,,
\end{equation}
is computed from the solution when the mode is much larger than the horizon and frozen in:
\begin{equation}
\Delta^2_\mathcal{R} (k) = \frac{k^3}{2\pi^2}\bigg|\frac{u_k}{z}\bigg|^2\,, \qquad \Delta^2_t(k) = \frac{4}{\pi^2}\frac{k^3}{M_{\rm pl}^2}\bigg|\frac{v_k}{a}\bigg|^2\,.
\end{equation}
As usual,  $\mathcal{R}$ denotes the curvature perturbations in comoving gauge,  and sometimes the notation $P_\mathcal{R}(k)= \frac{2\pi^2}{k^3}\Delta^2_\mathcal{R} (k)$ is used.

The amplitude of the power spectrum can also be computed in the slow roll approximation. 
Evaluated at $k_\ast$ (the scale which exits the horizon as $\phi = 0$), the amplitude is given in terms $H_\ast$ and the slow roll parameters by:
\begin{equation}
A_\mathrm{sr} = \frac{\big[1 - (2C + 1)\epsilon_\ast + C\eta_\ast\big]^2}{\pi\epsilon_\ast} \frac{H_\ast^2}{8\pi}\,,
\label{eq:Asr}
\end{equation}
where $A_\mathrm{sr}$ denotes $\Delta^2_\mathcal{R}(k_\ast)$ to second order in the slow-roll expansion. Here, $C = -2 + \ln 2 + \gamma$, and $\gamma$ is the Euler-Mascheroni constant \cite{Lidsey:1995np}.

\section{Method}\label{sec:method}

 Slow Roll Reconstruction treats the individual terms in the slow roll hierarchy as free parameters. By implementing the potential, equation~(\ref{eq:V}), within \ModeCode\ \cite{Mortonson:2010er,Easther:2011yq} we  obtain posterior distributions for $\{\epsilon_\ast,\eta_\ast,\xi_\ast\}$ (or  subsets of these variables) and compute Bayesian evidence to determine the optimal truncation-order for this expansion. \ModeCode\ modifies \CosmoMC\ \cite{Lewis:2002ah}, which is used in conjunction with the \MultiNest\ sampler \cite{Feroz:2007kg,Feroz:2008xx}.

\subsection{Bayesian model comparison using \MultiNest}\label{sec:evidenceDef}

Consider the  probability distribution for a set of parameters $\alpha$, given a model $M$ and the data $D$, denoted by $P(\alpha|D,M)$ \cite{Cox:1946}.   Bayes' theorem  yields the {\em  posterior\/}
\begin{equation}
P(\alpha|D,M) = \frac{P(D|\alpha,M)P(\alpha|M)}{P(D|M)}\,,
\end{equation}
where $\mathcal{L} \equiv P(D|\alpha,M)$ is the likelihood, $P(\alpha) \equiv P(\alpha|M)$ is the prior, and $E\equiv P(D|M)$ is the Bayesian evidence. Since the probability is normalized to unity we can compute the evidence directly, via
\begin{equation}
E = \int \mathrm{d}\alpha^N \mathcal{L}(\theta)P(\alpha)\,,
\label{eq:evidence}
\end{equation}
which is an $N$-dimensional integral  over a volume defined by the parameter ranges permitted by the prior.  

Using Bayes' theorem again, we have
\begin{equation}
\frac{P(M_1|D)}{P(M_2|D)} = \frac{E_1}{E_2}\frac{P(M_1)}{P(M_2)}\,.
\end{equation}
Given the {\em a priori\/} probability ratio $P(M_1)/P(M_2)$,  $E_1/E_2$ yields the ratio of probabilities for the two models, in the light of the data.   Any {\em a priori\/} preference for $M_1$ over $M_2$ is quantified by $P(M_1)/P(M_2)$; this ratio is always set to unity in our analysis.  Consequently, the evidence ratio $E_1/E_2$ expresses the relative ``betting odds" for  two models.

 The evidence, equation~(\ref{eq:evidence}), depends on both the value (or height, given that this is a positive definite integrand) of the likelihood $\mathcal{L}$ and the total volume of the parameter space.   The evidence thus encodes the  understanding  that a model in which $\mathcal{L}$  has a substantial amplitude over a large fraction of the permitted parameter volume is more predictive than one which does not \cite{Easther:2011yq}.

\subsection{Implementing Slow Roll Reconstruction}

We  fix $\phi = 0$  to correspond to the time when the pivot scale $k_\ast$  leaves the horizon and choose $k_\ast  = 0.05\;\mathrm{Mpc^{-1}}$.  We sample the slow roll parameters at the pivot scale, with priors  specified in Section \ref{sec:models}.  In principle, the initial value of the field $\phi_i$ is determined by the value of $\phi$ at the onset of inflation, or at which the density approaches Planckian values. In practice, the largest physical scale to which data are sensitive is $k_\mathrm{min} \sim 10^{-5}\;\mathrm{Mpc^{-1}}$ and we set $\phi_i$ by solving for the value of $\phi$ at which the mode with wavenumber $k=k_\mathrm{min}/100$ leaves the horizon.  The initial value of $\dot{\phi}_i$  is fixed using the slow roll attractor  $\dot{\phi}_i = -2M_\mathrm{pl}^2H'(\phi_i)$. We discard models for which inflation breaks down (\emph{i.e.} $\epsilon > 1$) as the field evolves from $\phi=\phi_i$ to $\phi = 0$, as these scenarios are not consistent with the assumption that the primordial universe is  inflating. 

The smallest scale we  consider is $k_\mathrm{max}$ which we take to be $k_\mathrm{max} \sim 10^2\;\mathrm{Mpc^{-1}}$, the smallest scale accessible to cosmological observations.  The power spectrum is  computed from the amplitudes of the solutions to the mode equations when the corresponding scale is far outside the horizon, so for self-consistency  we  require inflation to last until $k_\mathrm{max}$ is a hundred times larger than the horizon. Assuming $H$ to be roughly constant during inflation (which is consistent with limits from data), this puts a lower bound on the number of  $e$-folds of inflation  occurring after the pivot scale leaves the horizon, 
\begin{equation}
N_\mathrm{min} \gtrsim \log\bigg(\frac{100\, k_\mathrm{max}}{k_\ast}\bigg) \approx 12\,.
\end{equation}
We  discard models with fewer than $N_\mathrm{min}$ $e$-folds, measured from the instant  $k_\ast$ leaves the horizon. 

As implemented here, Slow Roll Reconstruction only considers the piece of the potential which is directly probed by cosmological data.   This amounts to marginalizing over the value of $\phi$ at which inflation ends -- physically, this can occur for example because the potential has a sharp ``cliff'' at some value of $\phi$ for which $\epsilon > 1$ when computed from equation~(\ref{eq:V}). Alternatively, this algorithm  can also be implemented by evolving $\phi$ until inflation naturally terminates ($\ddot{a} <0$)  and then imposing a prior based on the total number of $e$-foldings that occur after the pivot scale leaves the horizon \cite{Peiris:2008be}. By contrast, in this analysis we assume nothing about the evolution of the universe after $k_\mathrm{max}$ has reached its asymptotic value after leaving the horizon.

\subsection{Models and Priors}\label{sec:models}

The parameters which define the Hubble slow roll model to third order in the expansion defined in Eq.~\eqref{eq:SRRdef} are $H_\ast$, $\epsilon_\ast$, $\eta_\ast$ and $\xi_\ast$. But for all the models we consider, $H_\ast$  can, in principle, vary over many orders of magnitude and there is little clear {\em a priori\/} theoretical justification for restricting this range. However, given the slow roll parameters, we see from equation \eqref{eq:Asr} that $H_\ast$ is strongly correlated with the amplitude of the primordial spectrum.
Consequently, we treat $\Delta^2_\mathcal{R}(k_\ast)$,  the amplitude of the primordial scalar power spectrum, for which the choice of prior range is more straightforward, as an independent parameter with a logarithmic prior. 
We allow a range that  is generous relative to current constraints on this parameter (e.g. Ref. \cite{Komatsu:2010fb}). However our results do not depend significantly on the specific choice, and setting a  narrower range in the prior would implicitly inject information from the data  used to compute the posterior. We obtain $H_\ast$ directly from equation \eqref{eq:Asr}, given values for $A_\mathrm{sr}$ and the slow roll parameters. This is the one point at which we use the slow roll {\em approximation}; the precise amplitude of the power spectrum at the pivot $\Delta^2_\mathcal{R}(k_\ast)$ extracted from a numerical solution of the mode equations  differs slightly from  $A_\mathrm{sr}$. In principle we could solve numerically for the exact value of $H_\ast$ needed to reproduce $\Delta^2_\mathcal{R}(k_\ast)$, but $H_\ast$ only changes at the level of a few percent if this is done. Note that the slow roll approximation is used only in order to set this prior; for the likelihood computation the numerical solution for $\Delta^2_\mathcal{R}(k_\ast)$ is used.

The specific set of models we consider is:
\begin{itemize}
\item ``$\epsilon_\ast$ only'':  $\eta_\ast = \xi_\ast = 0$,  $\epsilon_\ast$ and $A_\mathrm{sr}$ are varied.
\item ``$\eta_\ast$ only'': $\epsilon_\ast = 10^{-6}$, $\xi_\ast = 0$,  $\eta_\ast$ and $A_\mathrm{sr}$ are varied.
\item ``$\epsilon_\ast$, $\eta_\ast$'': $\xi_\ast = 0$,  $\epsilon_\ast$, $\eta_\ast$ and $A_\mathrm{sr}$ are varied.
\item ``$\epsilon_\ast$, $\eta_\ast$, $\xi_\ast$'': all parameters varied.
\end{itemize}
The ``$\eta_\ast$ only'' case is the ``Low-$\epsilon$, 1-Parameter'' scenario described in  Ref. \cite{Adshead:2008vn}: When $\epsilon$ is much smaller than $1$, it effectively  decouples from the slow roll hierarchy, meaning that the inflationary dynamics are then independent of the exact value of $\epsilon$ and are determined by the other slow roll parameters, $\eta$ in this case. We implement this by fixing $\epsilon_\ast$ to a small, finite value and have checked that the results we obtain are insensitive to the specific choice at which we fix $\epsilon_\ast$. 

We consider both uniform and logarithmic priors on $\epsilon_\ast$, and impose uniform priors on $\eta_\ast$ and $\xi_\ast$.  Since the tensor-to-scalar ratio is strongly correlated with $\epsilon$,  a uniform prior for this parameter biases the analysis toward models where inflation occurs at high energy scales,  which are likely to produce a significant background of tensor perturbations \cite{Valkenburg:2008cz}.  Likewise,  $A_\mathrm{sr}$ is drawn from a logarithmic prior.  The prior ranges for all model parameters are listed in Table \ref{tab:ranges}, while the priors for other free cosmological parameters are specified in Table \ref{tab:cosmoRanges}.

\begin{table}
\begin{tabular}{|c|c|c|}
\hline
Parameter & Lower limit & Upper limit
\\ \hline \hline
$\log_{10} \epsilon_\ast$ & $-10$ & $0$ 
\\ \hline
$\epsilon_\ast$ & $0$ & $1$
\\ \hline
$\eta_\ast$ & $-1$ & $1$ 
\\ \hline
$\xi_\ast$ & $-1$ & $1$ 
\\ \hline
$\log(10^{10} A_\mathrm{sr})$ & $2.7$ & $4$
\\ \hline
\end{tabular}
\caption{Prior ranges for the model parameters. Runs are performed with both  log and uniform priors on $\epsilon_\ast$; the corresponding ranges are given in the first two rows.}
\label{tab:ranges}
\end{table}
\begin{table}
\begin{center}
\begin{tabular}{|l|c|}
\hline
Parameter &   Prior   \\
\hline \hline
Baryon fraction  & $0.015 < \Omega_b h^2 < 0.035$ \\
 \hline
Dark matter  &  $0.05 < \Omega_{\rm dm} h^2 < 0.2$  \\
\hline
Reionization & $0.01 < \tau <  0.25$ \\
\hline
Projected acoustic scale & $0.8 <\theta < 1.2 $ \\
\hline
Sunyaev-Zel'dovich Amplitude &  $0<A_{\rm SZ} < 2$ \\
\hline

\end{tabular}
\end{center}
\caption{Prior ranges for other free parameters. All priors are uniform. The universe is assumed to be flat ($\Omega_k = 0$), so that the fractional energy density of dark energy is not an independent parameter. The effective number of relativistic species is set to 3.046. For the SPT dataset, we also include two nuisance parameters for the power from Poisson and clustered point sources (see \cite{Keisler:2011aw} for details).}
\label{tab:cosmoRanges}
\end{table}

\begin{figure}[tb]
\includegraphics[width=0.48\textwidth]{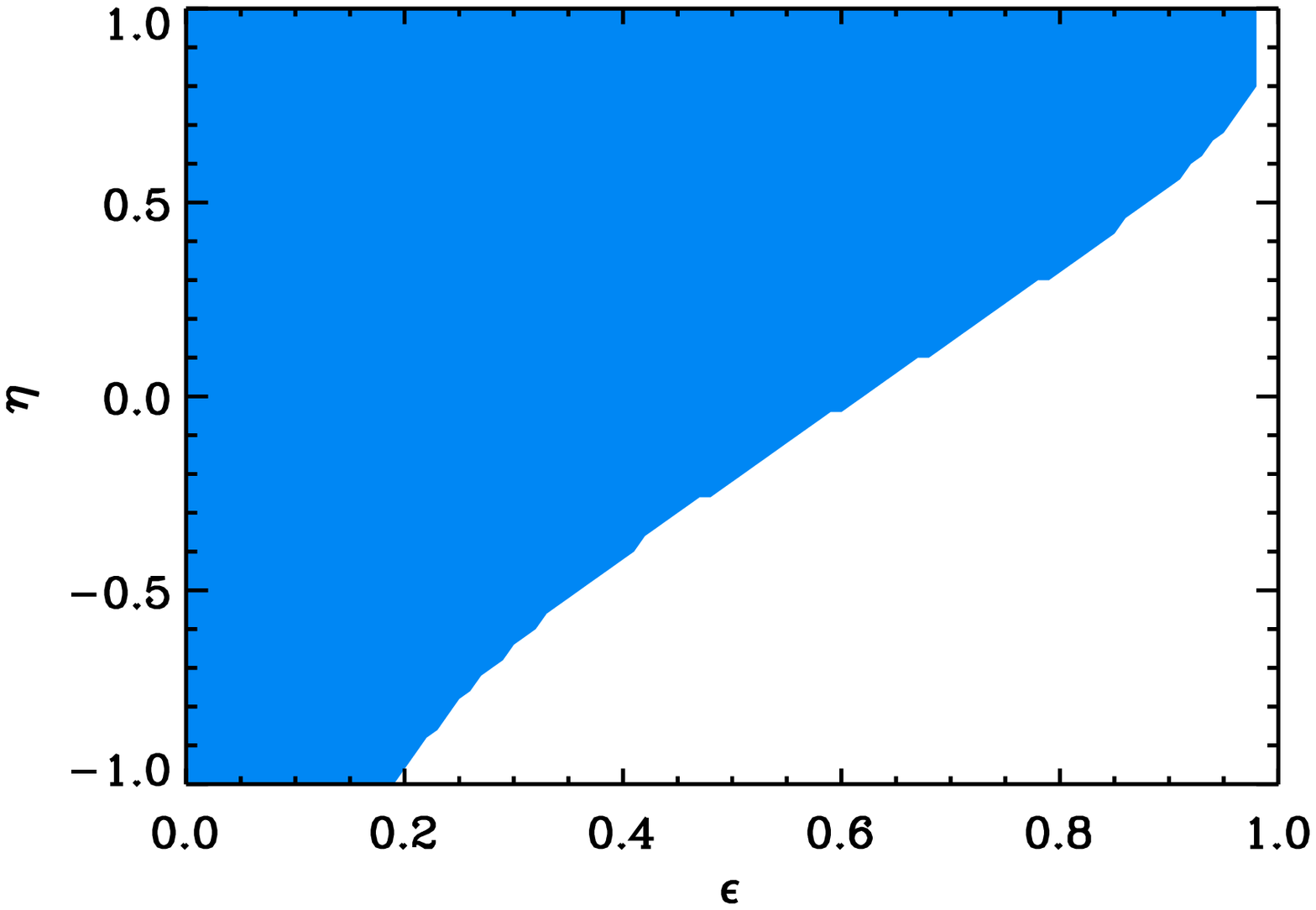}
\includegraphics[width=0.48\textwidth]{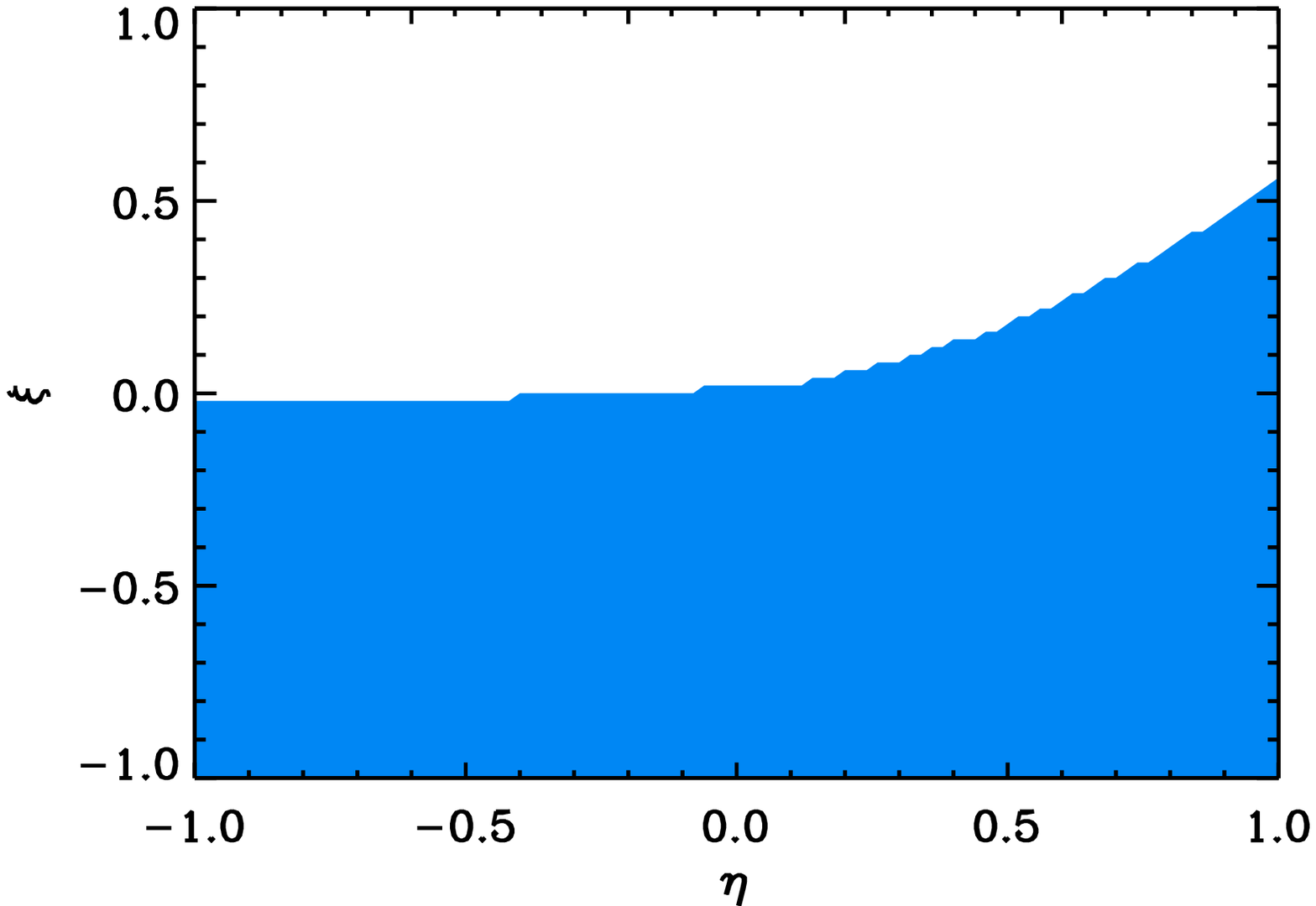}
\caption{The shaded region shows the projections of the prior volume after imposing the condition that the model achieves at least $N_{\mathrm{min}} = 12$ $e$-folds of inflation after the pivot scale leaves the horizon.}
\label{fig:priors}
\end{figure}

The ranges of the slow roll parameters allowed by these priors are substantially larger than those used in most previous analyses \cite{Peiris:2006sj,Peiris:2006ug,Peiris:2008be}; since we are computing the evidence we only want to stipulate that inflation is taking place, without requiring that the slow roll parameters are small, in order to compute a self-consistent value for the evidence.    Requiring that inflation lasts $N_{\mathrm{min}}$ $e$-folds after the pivot scale exits the horizon imposes an additional prior on the slow roll parameters. The resulting joint prior on any pair of slow roll parameters is no longer ``rectangular'' -- Figure \ref{fig:priors} illustrates  the effective priors  after this constraint is imposed.  

 If $N_{\mathrm{min}}$ is increased, the allowed region of parameter space   becomes smaller, as we discuss  in Section \ref{sec:evidence}.  A small volume of parameter space for which inflation lasts longer than $N_{\mathrm{min}}$ $e$-folds  is excluded by our prior ranges on the slow roll parameters -- however, this region is small and we have checked that modifying this choice does not impact our results.

\subsection{Data}\label{sec:data}

\begin{figure}[tb]
\includegraphics[width=0.48\textwidth]{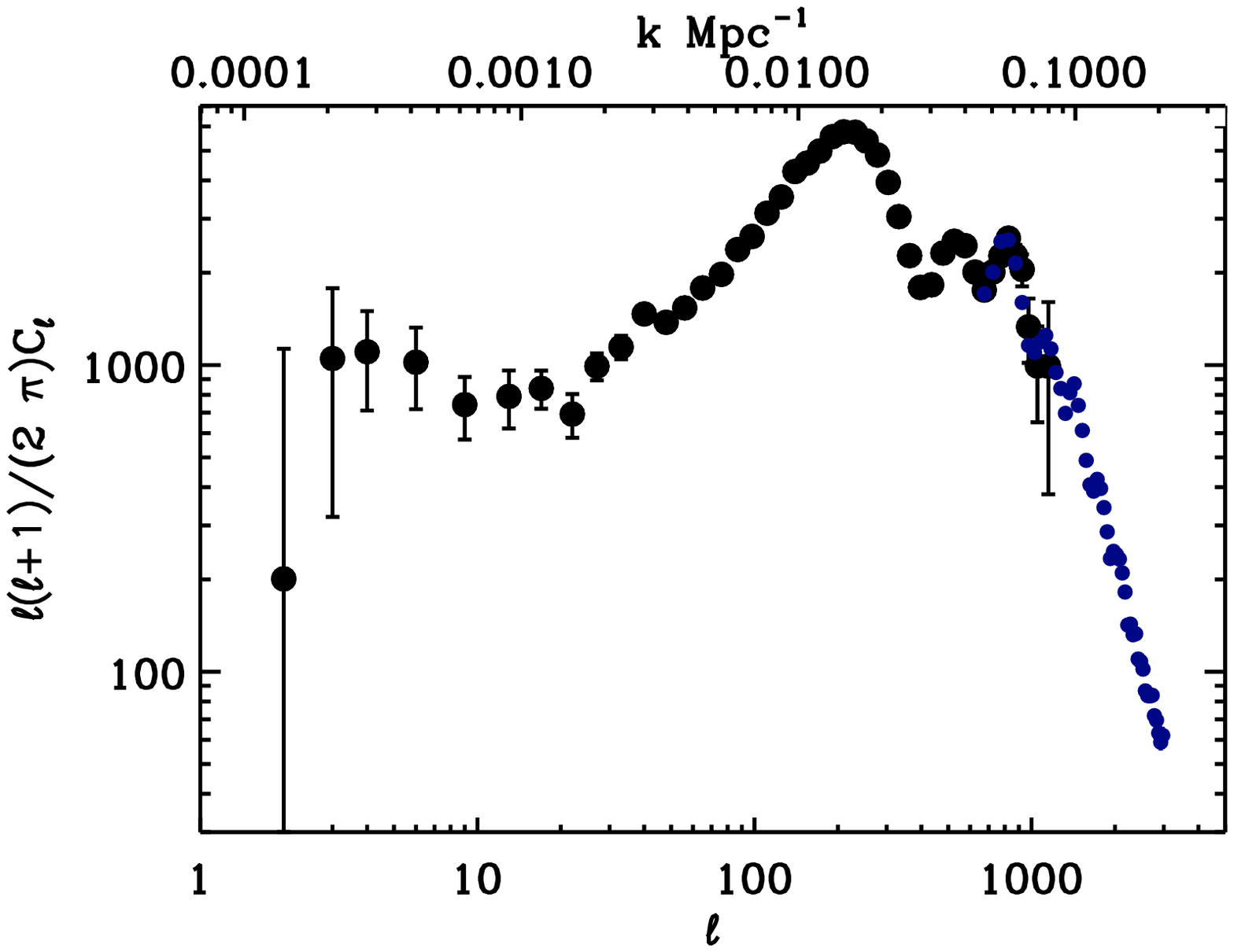}
\caption{The CMB temperature angular power spectrum as measured by WMAP~7 year data (black dots) and the South Pole Telescope (blue dots). The error bars on the SPT data are smaller than the dots.}
\label{fig:cmbdata}
\includegraphics[width=0.48\textwidth]{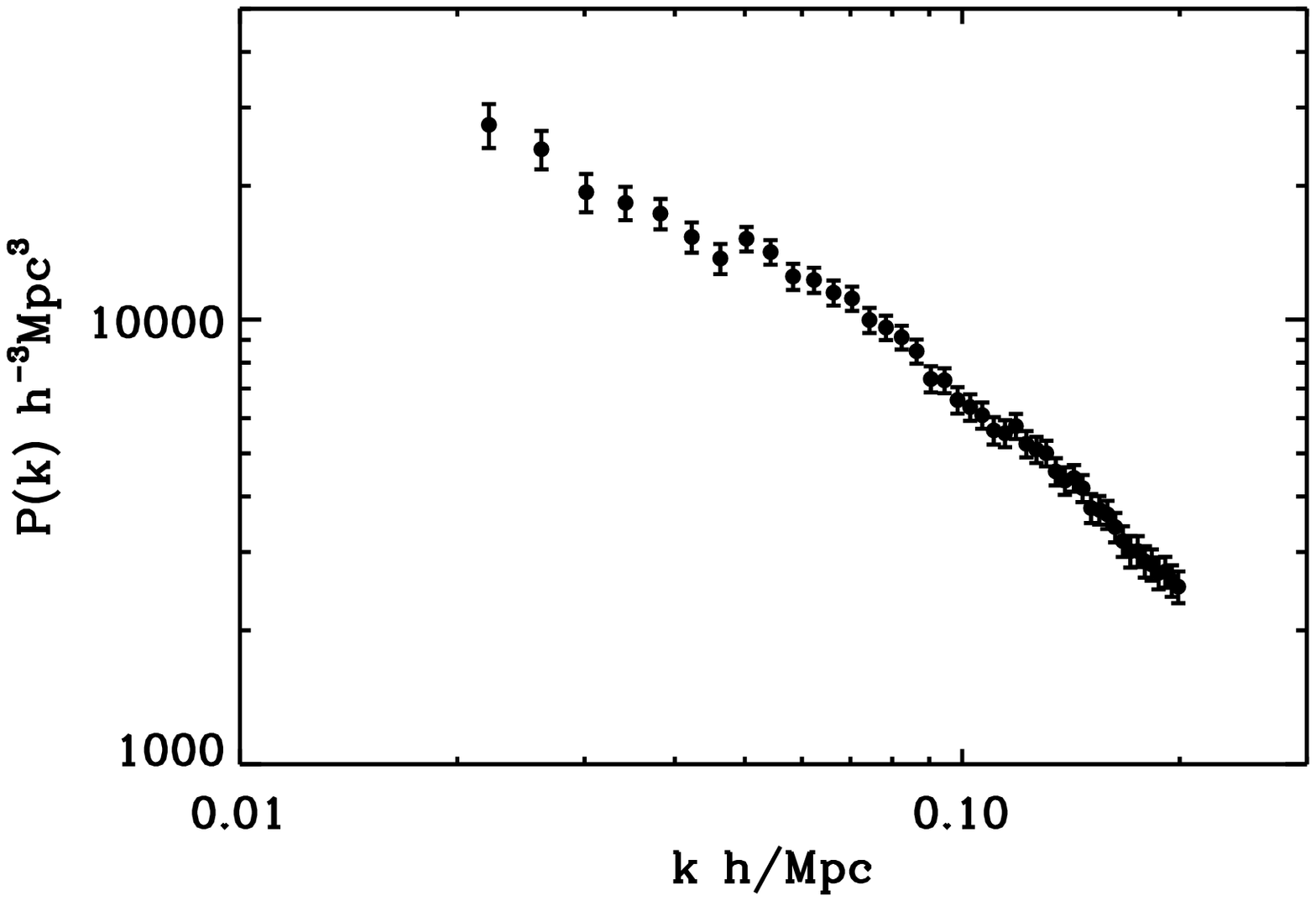}
\caption{LRG power spectrum from the SDSS data release 7.}
\label{fig:lrgdata}
\end{figure}

We use the WMAP~7 year likelihood (WMAP7) \cite{Larson:2010gs} with data on the damping tail of the CMB temperature power spectrum from the South Pole Telescope \cite{Keisler:2011aw}.  We follow Ref. \cite{Keisler:2011aw} in computing the SPT likelihood,  setting $l_{\mathrm{max}} = 3000$,  and marginalizing over foreground contributions from unresolved point sources and Sunyaev-Zel'dovich (SZ) clusters. These data are presented in Figure \ref{fig:cmbdata}, which also shows the physical wavenumbers which contribute to the corresponding angular scales.  In addition, we use the power spectrum of luminous red galaxies (LRG) derived from the SDSS data release 7 \cite{Reid:2009xm}.  We adopt the public LRG likelihood code released by the SDSS collaboration, marginalizing over the amplitude and using only the information from the shape of the power spectrum. This likelihood automatically marginalizes over a possible scale dependence of the small-scale bias. The LRG power spectrum is shown in Figure \ref{fig:lrgdata}.

In what follows we use the data combinations WMAP, WMAP+SPT and WMAP+SPT+LRG -- the dependence of the evidence ratios on the different data combinations reveals the information content of each dataset in this model selection problem.

\section{Results}\label{sec:results}

\subsection{Parameter estimation}\label{sec:constraints}

\begin{figure}[tb]
\includegraphics[width=0.48\textwidth]{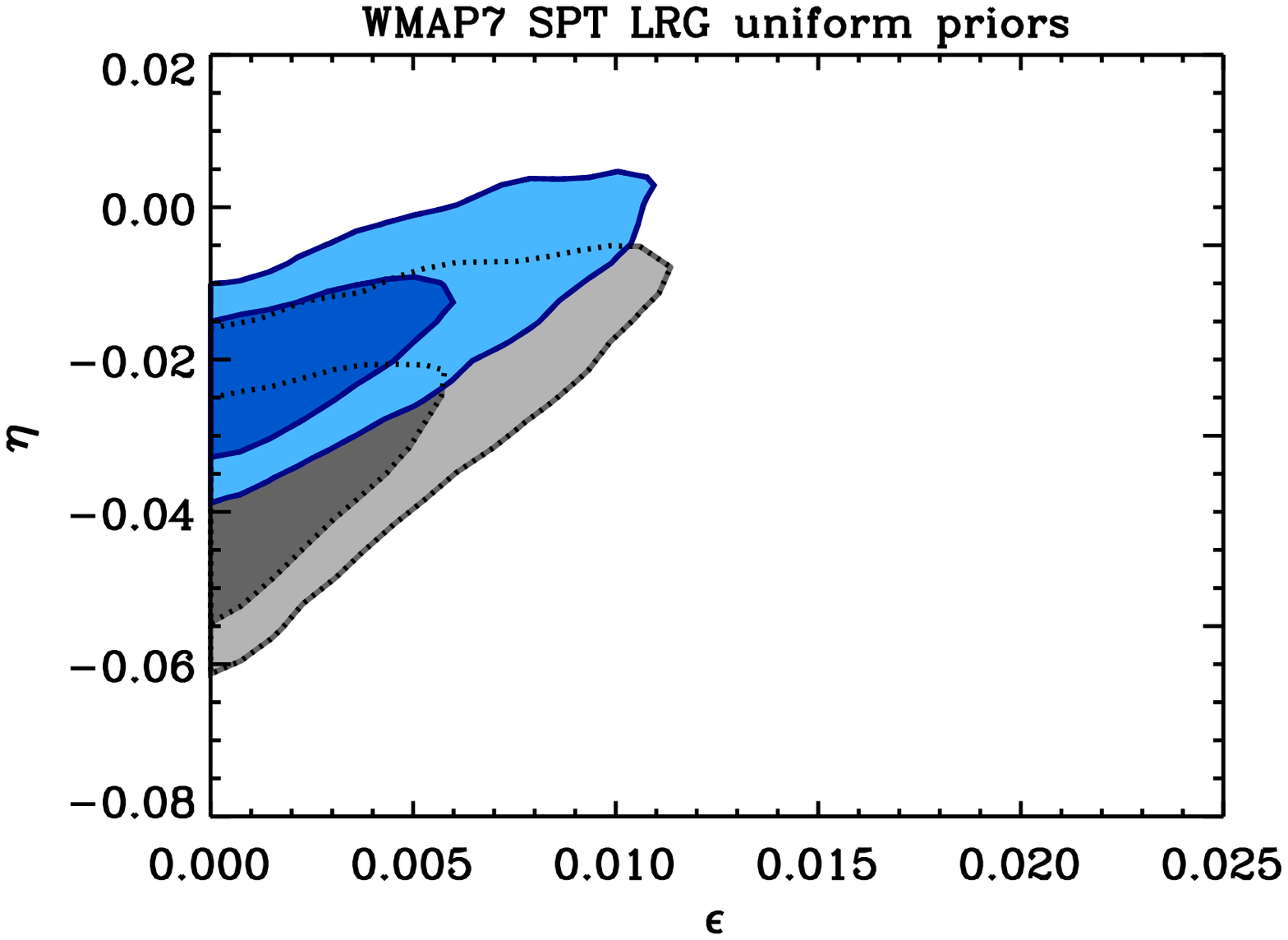}
\includegraphics[width=0.48\textwidth]{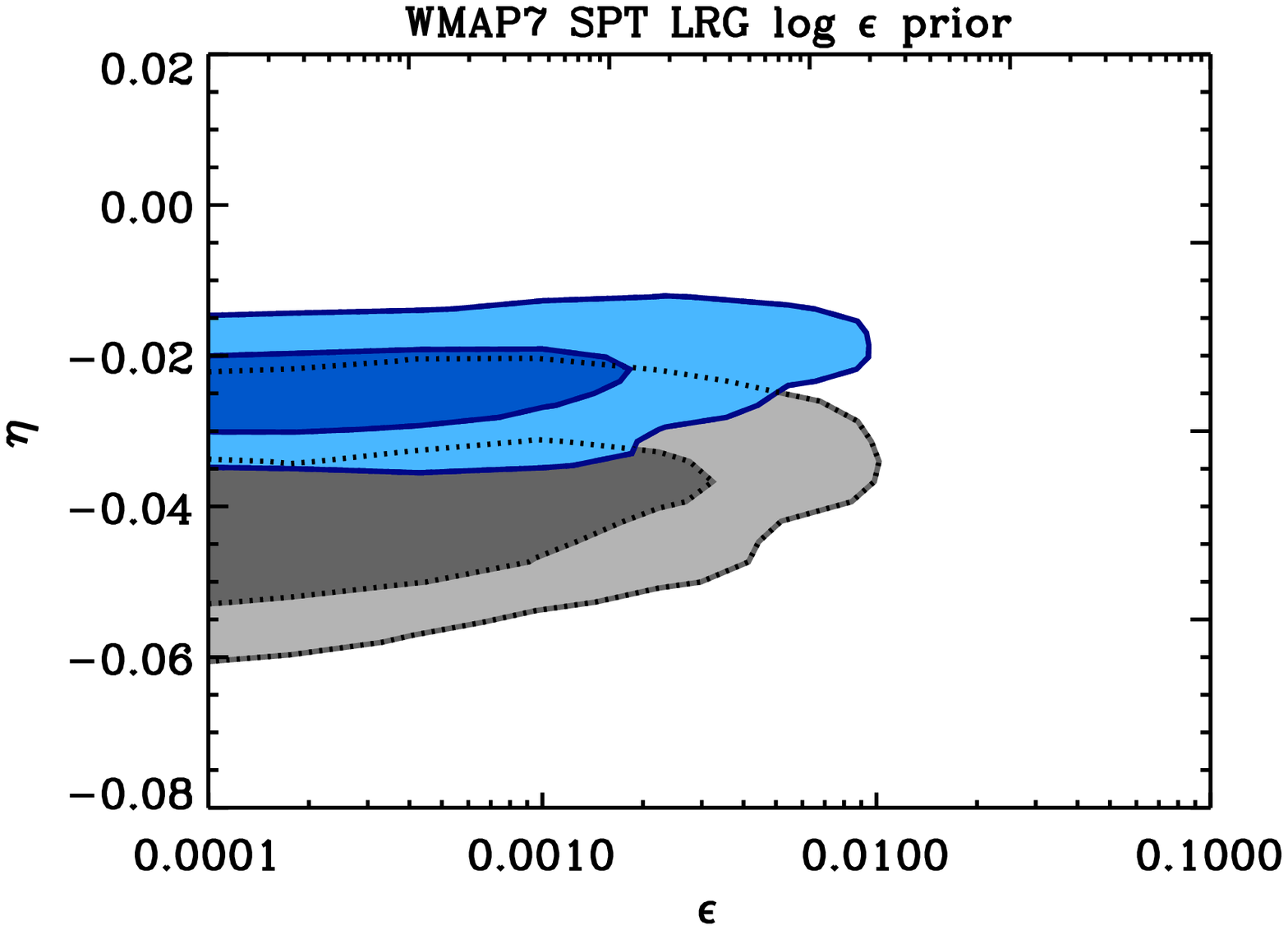}
\caption{Joint constraints on $\epsilon_\ast$ and $\eta_\ast$  using data from WMAP7, SPT and SDSS LRGs,  marginalizing over all other parameters. The blue (upper) contours correspond to the ``$\epsilon_\ast$, $\eta_\ast$'' model while the grey (lower) contours correspond to the ``$\epsilon_\ast$, $\eta_\ast$, $\xi_\ast$'' model. Contours correspond to $68$ and $ 95$\% joint confidence levels; blue denotes the ``$\epsilon_\ast$, $\eta_\ast$'' model and grey denotes the `$\epsilon_\ast$, $\eta_\ast$, $\xi_\ast$'' model. Top: uniform prior on $\epsilon_\ast$. Bottom: log prior on $\epsilon_\ast$.}
\label{plot:Constraints}
\end{figure}

\begin{figure}[tb]
\includegraphics[width=0.47\textwidth]{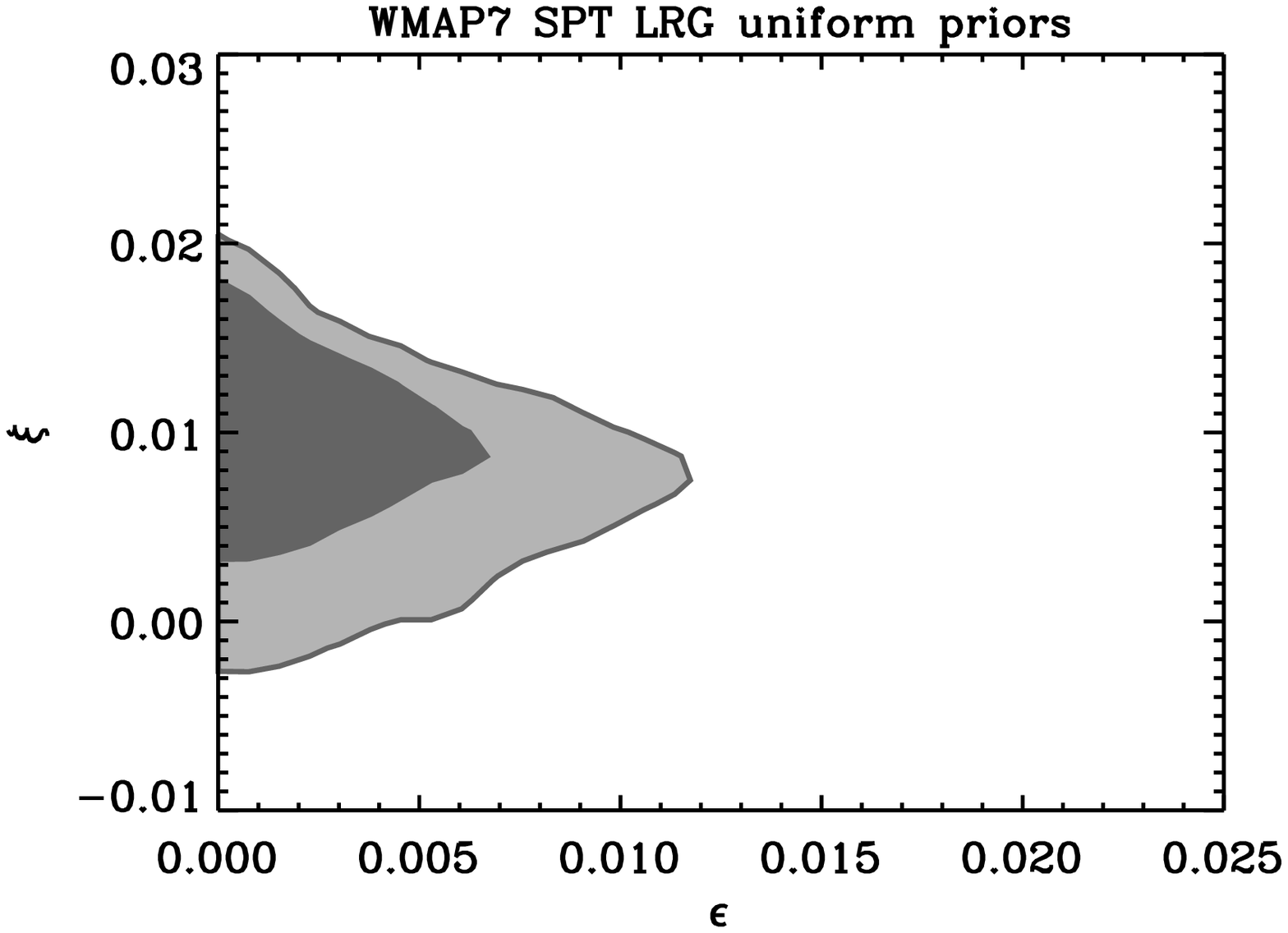}
\includegraphics[width=0.47\textwidth]{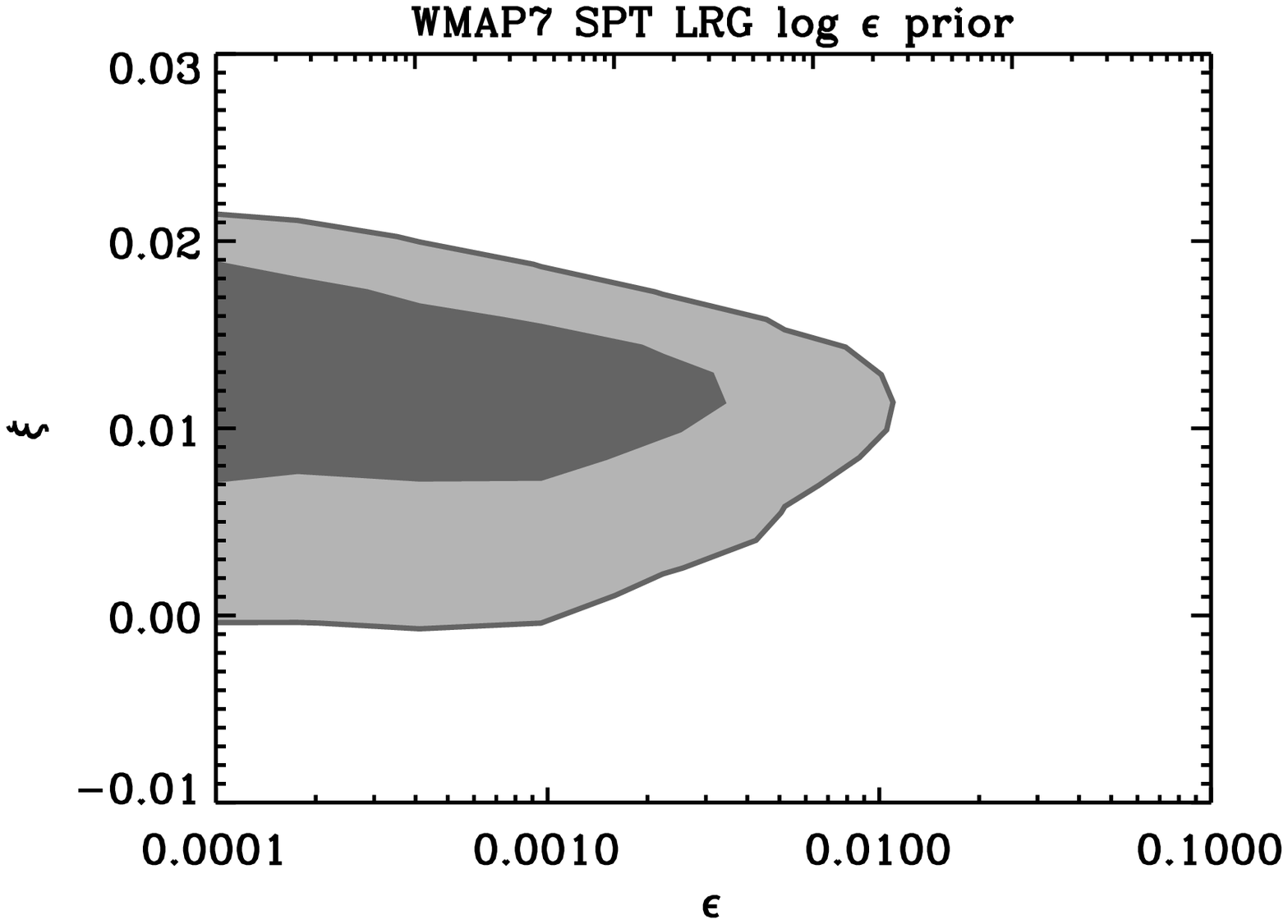}
\caption{Joint constraints on $\epsilon_\ast$ and $\xi_\ast$ using data from WMAP7, SPT and SDSS LRGs,  marginalizing over all other parameters. Contours correspond to $68$ and $ 95$\% joint confidence levels. Top: uniform prior on $\epsilon_\ast$. Bottom: log prior on $\epsilon_\ast$.}
\label{plot:xiConstraints}
\end{figure}

\begin{figure}[tb]
\includegraphics[width=0.47\textwidth]{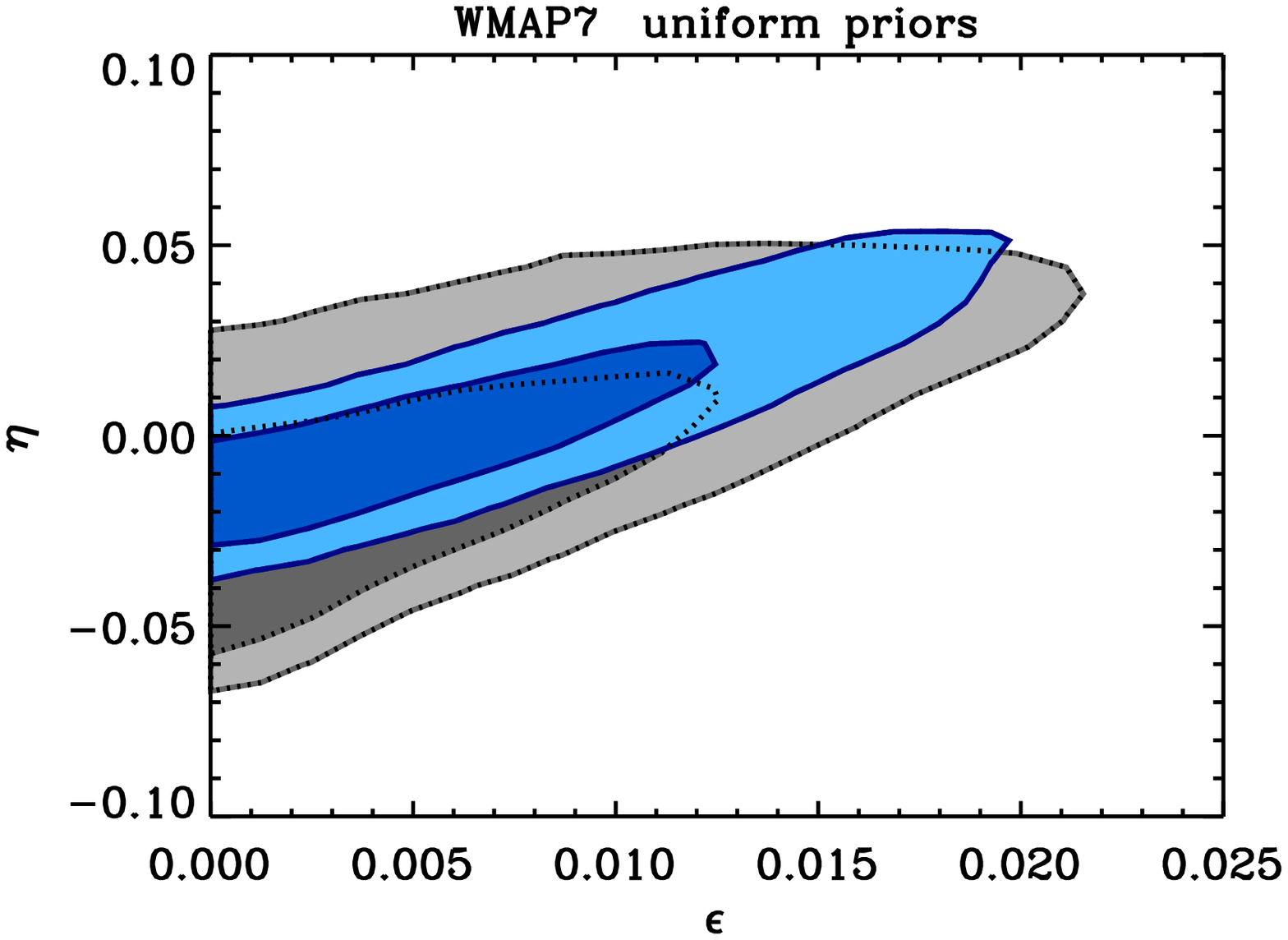}
\includegraphics[width=0.47\textwidth]{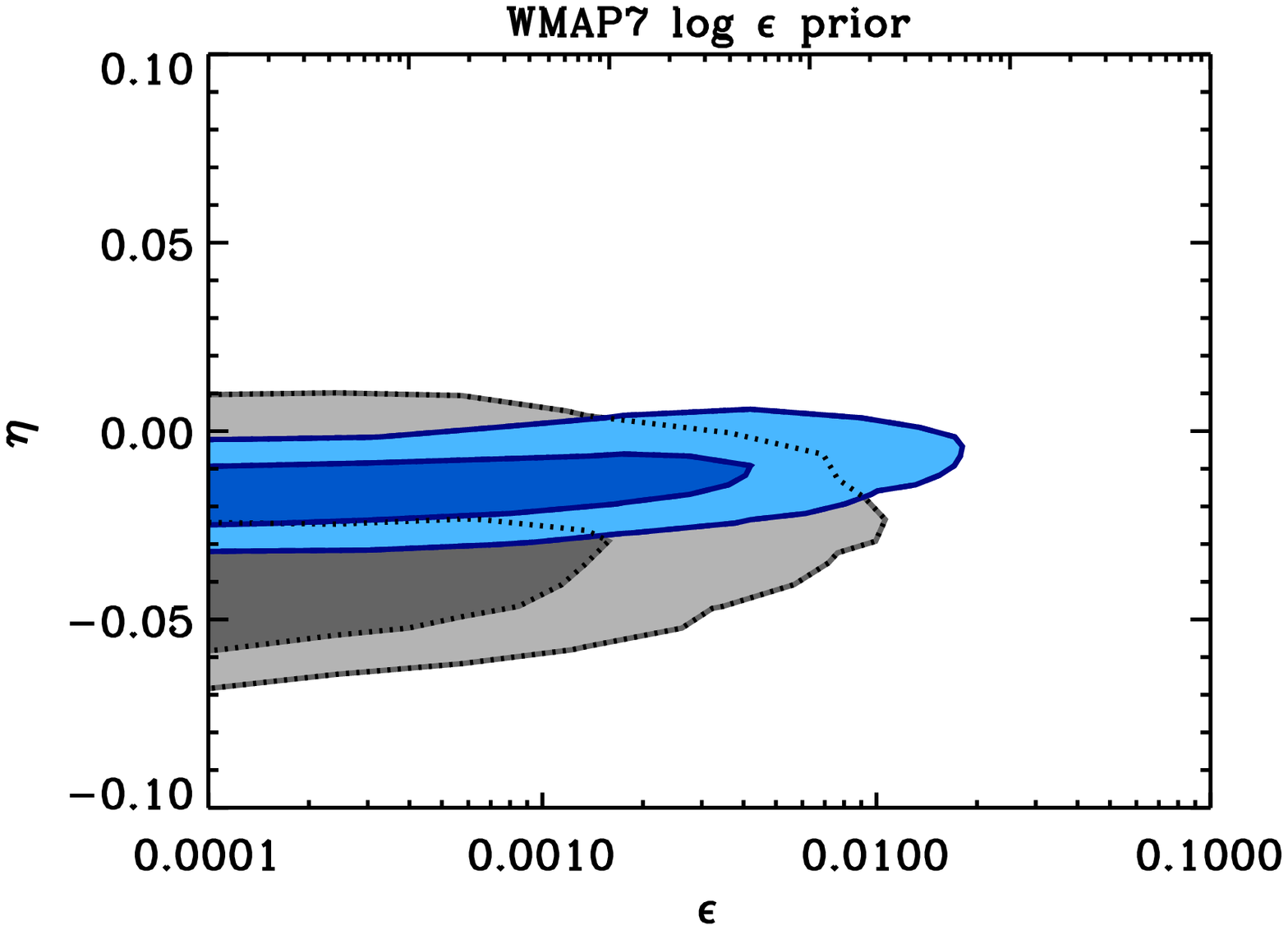}
\caption{Joint constraints on $\epsilon_\ast$ and $\eta_\ast$ using data from WMAP7 only, after marginalizing over all other parameters. The contours are in the same format as in Figure \ref{plot:Constraints}. Top: uniform prior on $\epsilon_\ast$. Bottom: log prior on $\epsilon_\ast$.}
\label{plot:wmapConstraints}
\end{figure}

Figure \ref{plot:Constraints} presents the posterior constraints from WMAP+SPT+LRG on $\epsilon_\ast$ and $\eta_\ast$ for  the ``$\epsilon_\ast$, $\eta_\ast$''  and ``$\epsilon_\ast$, $\eta_\ast$, $\xi_\ast$'' models, while Figure \ref{plot:xiConstraints} show the constraints on $\epsilon_\ast$ and $\xi_\ast$ for both logarithmic and uniform priors on~$\epsilon_\ast$. Appendix \ref{sec:bounds} tabulates the constraints on each of the slow roll parameters for all combinations of model, $\epsilon$ prior and dataset. 

Observe that the posteriors all peak in regions of parameter space where the slow roll parameters are small. Consequently we can conclude that, within the context of this model, slow roll inflation is preferred by the data. Constraints on $\eta_\ast$ differ when the uniform prior on $\epsilon_\ast$ is replaced with a logarithmic prior. We interpret this as the data currently not being informative enough to overcome the priors; see section \ref{sec:evidence} for further discussion. 

For comparison, we show the same information in Figure \ref{plot:wmapConstraints} using only WMAP7 data. We see that the addition of SPT and LRG data lead to better constraints on $\eta_\ast$. 
Note that one cannot directly compare the constraint $\epsilon_\ast<0.016$ (95\% CL) to the constraint on the tensor-to-scalar ratio $r<0.36$ (95\% CL) reported by the WMAP7 team \cite{Larson:2010gs}. We find $r<0.27$ (95\% CL), which is close to the value one would obtain by simply using the slow roll approximation $r\approx 16 \epsilon_\ast$. This $\sim 30\%$ difference can be explained by the different choice for the pivot scale (Ref.~\cite{Larson:2010gs} uses $k_p = 0.002\;\mathrm{Mpc}^{-1}$) and the difference in the models used. The Hubble slow roll model ``$\epsilon_\ast$, $\eta_\ast$'' is not just a simple reparametrization of the phenomenological model parametrized by the spectral index $n_s$ and the tensor-to-scalar ratio $r$, which was applied in Ref.~\cite{Larson:2010gs}.

\subsection{Bayesian evidence}\label{sec:evidence}

\begin{figure*}
\includegraphics[width=0.7\textwidth,height=0.25\textheight]{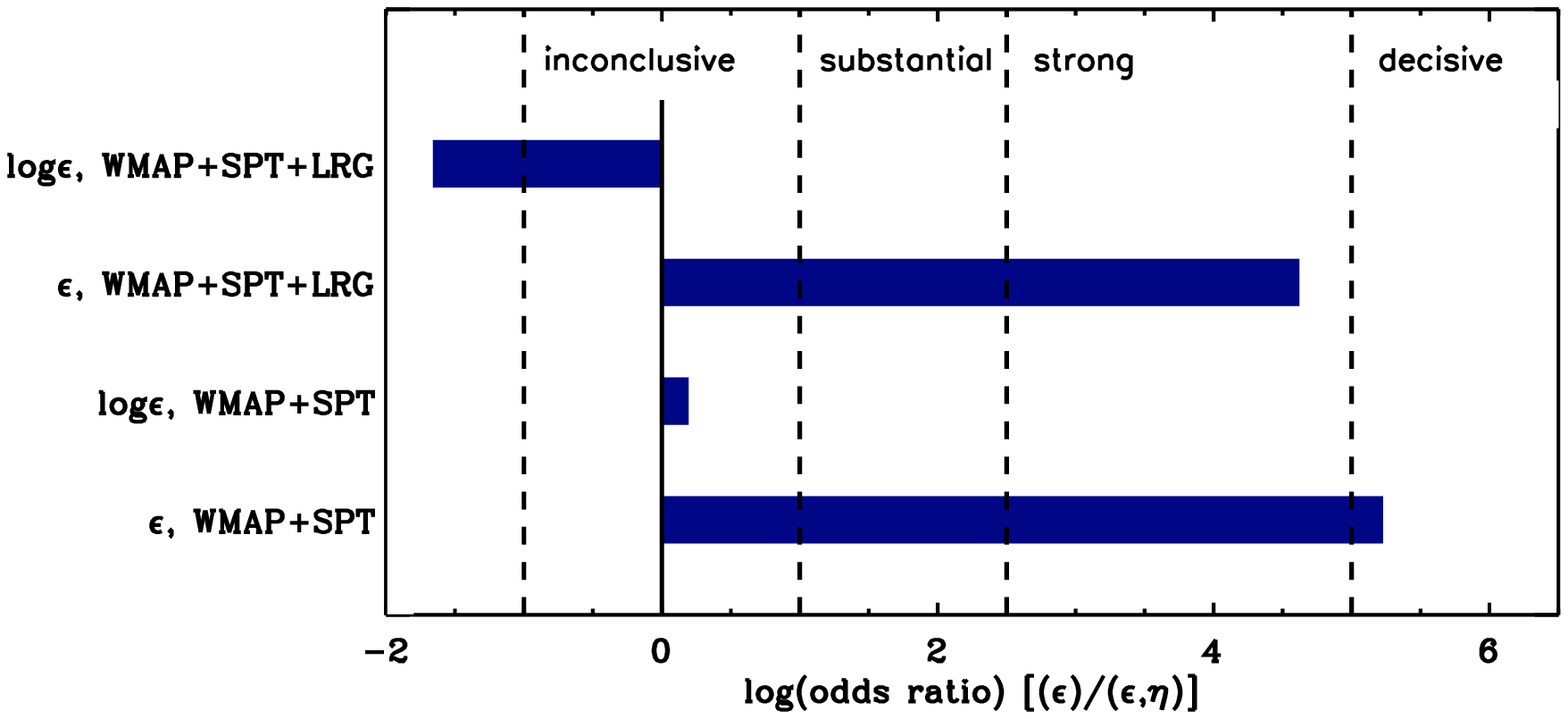}
\caption{$\Delta \log E(\epsilon_\ast;\epsilon_\ast, \eta_\ast)$ for log and uniform priors on $\epsilon_\ast$ and different data sets, as indicated on the vertical axis. The estimates of $\Delta \log E(\epsilon_\ast;\epsilon_\ast, \eta_\ast)$ have an uncertainty of $\sim 0.3$ at the numerical settings used in this work. When taking the ratios, the same priors were used in the numerator and denominator.}
\label{plot:epsiEvidence}
\end{figure*}

\begin{figure*}
\includegraphics[width=0.7\textwidth,height=0.25\textheight]{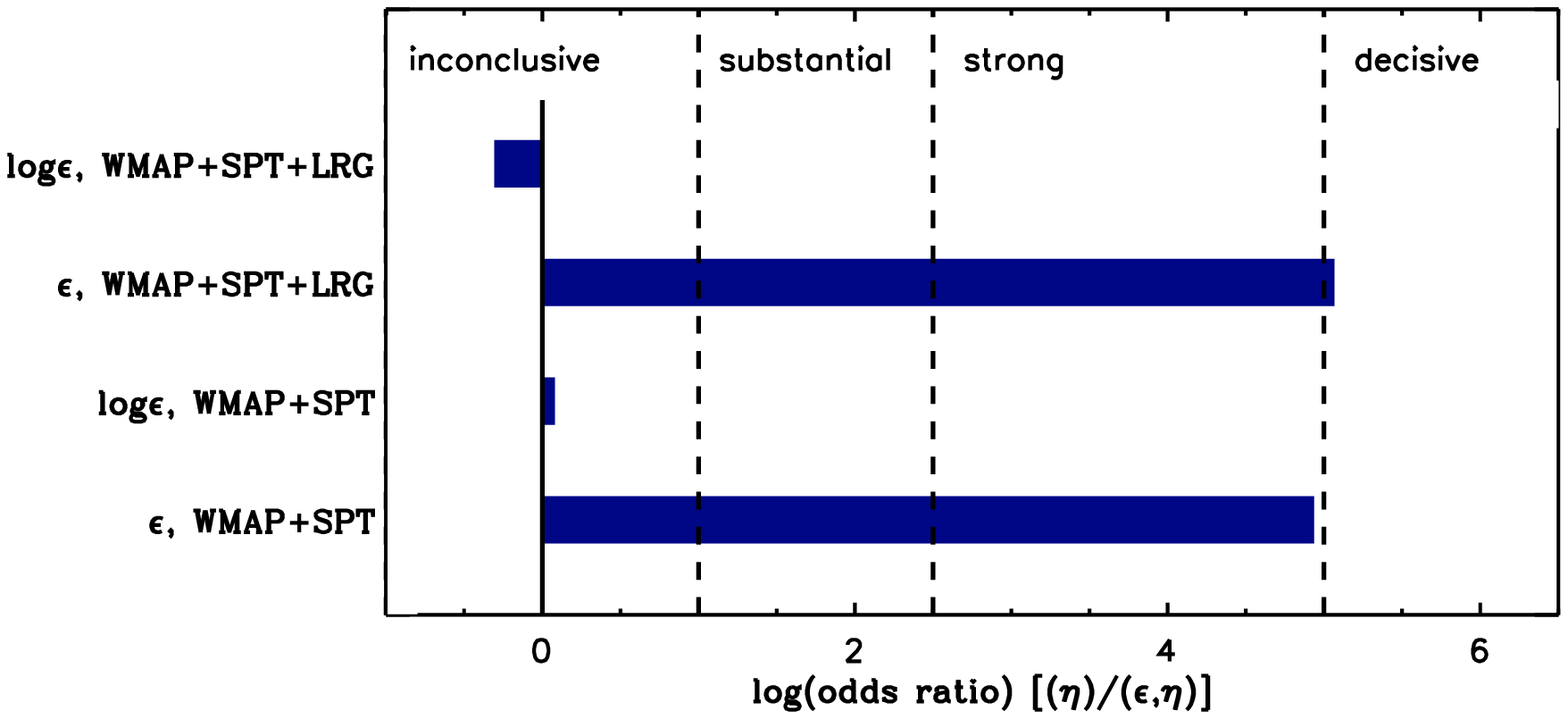}
\caption{$\Delta \log E(\eta_\ast;\epsilon_\ast, \eta_\ast)$ for log and uniform priors on $\epsilon_\ast$ and different data sets, as indicated on the vertical axis. The estimates of $\Delta \log E(\eta_\ast;\epsilon_\ast, \eta_\ast)$ have an uncertainly of $\sim 0.3$ at the numerical settings used in this work. When taking the ratios, the same priors were used in the numerator and denominator.}
\label{plot:etaEvidence}
\end{figure*}

As discussed in Section \ref{sec:evidenceDef}  the evidence is not an absolute quantity: the ratio of the evidence values for two models expresses the relative ``betting odds''  that  these  models are  responsible for the observed state of the universe. 
 In Figure \ref{plot:epsiEvidence} we plot $\Delta\log E(\epsilon_\ast;\epsilon_\ast,\eta_\ast) \equiv \log[E(\epsilon_\ast)/E(\epsilon_\ast, \eta_\ast)]$  for uniform and log priors on $\epsilon_\ast$. There appears to be ``strong'' evidence for the ``$\epsilon_\ast$ only'' model compared to the ``$\epsilon_\ast$, $\eta_\ast$'' model when using uniform priors, while the log priors give ``inconclusive'' results. A similar result is seen in Figure \ref{plot:etaEvidence}, which shows $\Delta\log E(\eta_\ast;\epsilon_\ast,\eta_\ast)$. Again the uniform priors on $\epsilon_\ast$ gives ``strong'' evidence for ``$\eta_\ast$ only'' scenario compared to the ``$\epsilon_\ast$, $\eta_\ast$'' model, while the evidence ratio is ``inconclusive'' with the log prior. 
   
In the cases with uniform $\epsilon_\ast$ priors, the evidence clearly tells us that a single parameter (either ``$\epsilon_\ast$ only'' or ``$\eta_\ast$ only'') is preferred by the data over the two-parameter ``$\epsilon_\ast$, $\eta_\ast$'' case; however, the current data are not strong enough to tell us \emph{which} of the single parameter models to pick.

When using the log $\epsilon_\ast$ priors, the parameter space is heavily weighted towards small $\epsilon_\ast$, approaching the ``Low-$\epsilon$" scenario, $\epsilon\ll 1$, where this parameter effectively decouples from the  Hubble slow roll hierarchy. Hence, in the scenarios with log priors, even though at face value we are comparing a two parameter model with a single parameter model, we are effectively comparing two single-parameter models. In the absence of a detection of tensor modes or a much stronger upper limit on tensors than is currently available, $\epsilon$ becomes a nuisance parameter which does not have a significant effect on the likelihood. 

We can gain further insight into these results by mapping the slow roll parameter into the usual spectral index $n_s$ and tensor-to-scalar ratio $r$ which, to first order in slow roll, 
\begin{align}
n_s - 1 &\approx 2\eta_\ast - 4\epsilon_\ast\,, \label{eq:ns}\\
r &\approx 16\epsilon_\ast\,. \label{eq:r}
\end{align}
Varying $\epsilon_\ast$ changes both $n_s$ and $r$, while varying $\eta_\ast$ changes only $n_s - 1$. Data favor models with $n_s - 1 \approx -0.05$. With a log prior the ``$\epsilon_\ast$ only'' model gives greater weight to the $r \approx 0$, $n_s \approx 1$ region of parameter space, which is disfavored by data,  compared to the uniform prior.   Further, with a log prior the likelihood changes very slowly over most of the range spanned by $\log{\epsilon_\ast}$. In this case, we would expect the evidence computed for the ``$\epsilon_\ast$, $\eta_\ast$''  model to approach the ``$\eta_\ast$ only''  model, as seen in Figure \ref{plot:epsiEvidence} and \ref{plot:etaEvidence}.   This result also mirrors earlier discussions of the prior dependence of estimates of the tensor-to-scalar ratio \cite{Valkenburg:2008cz}.

Considered together, Figs. \ref{plot:epsiEvidence} and \ref{plot:etaEvidence} for a uniform $\epsilon_\ast$ prior indicate that  the evidence ``decisively" prefers  a single-parameter model (to explain $n_s<1$) over a two-parameter model, but without a tensor detection (or a strong upper limit) it cannot discriminate whether that parameter should be $\epsilon$ or $\eta$.

\begin{figure*}
\includegraphics[width=0.7\textwidth]{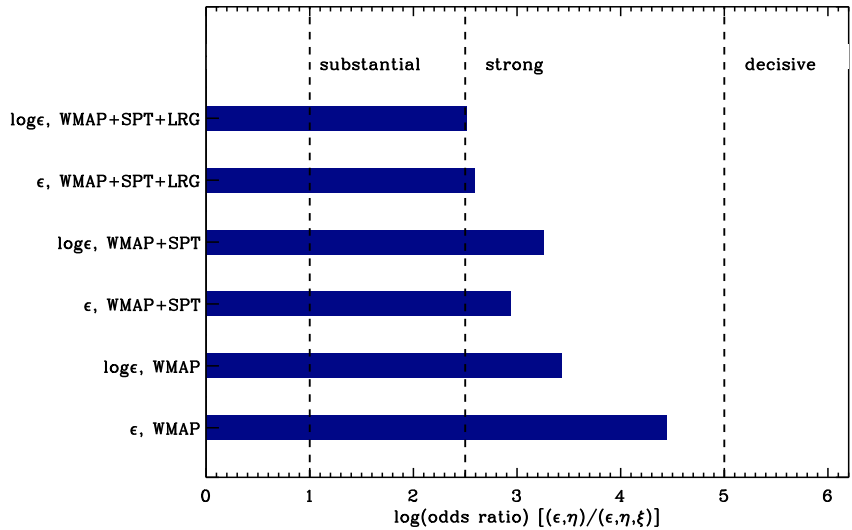}
\caption{$\Delta \log E(\epsilon_\ast,\eta_\ast;\epsilon_\ast, \eta_\ast,\xi_\ast)$ for log and uniform priors on $\epsilon_\ast$ and different data sets as indicated in the vertical axis. The values of $\Delta \log E(\epsilon_\ast,\eta_\ast;\epsilon_\ast, \eta_\ast,\xi_\ast)$ have an uncertainty of $\sim 0.3$ at the numerical settings used in this work.}
\label{plot:Evidence}
\end{figure*}

In Figure \ref{plot:Evidence} we plot  $\Delta \log E(\epsilon_\ast,\eta_\ast; \epsilon_\ast,\eta_\ast,\xi_\ast)$ for different choices of $\epsilon$ priors and data. According to the usual scale \cite{Jeffreys} for comparing models, there is ``strong'' evidence in favour of the ``$\epsilon_\ast$, $\eta_\ast$'' model compared to the ``$\epsilon_\ast$, $\eta_\ast$, $\xi_\ast$'' model, indicating that  $\xi_\ast$ is not needed to explain the data, and this result is largely independent of the prior on $\epsilon_\ast$.  However, adding SPT and LRG data \emph{decreases} the significance of this conclusion. If the inflationary phase was well described by only $\epsilon_\ast$ and $\eta_\ast$, adding more data should increase the evidence ratio, in  the absence of systematic effects in the data. However, even if the true underlying model  had only $\epsilon_\ast$ and $\eta_\ast$, a small systematic mismatch between different data sets (e.g. normalization issues) can lead to a spurious preference for a non-zero~$\xi_\ast$.

\begin{figure}[ht]
\includegraphics[width=0.48\textwidth]{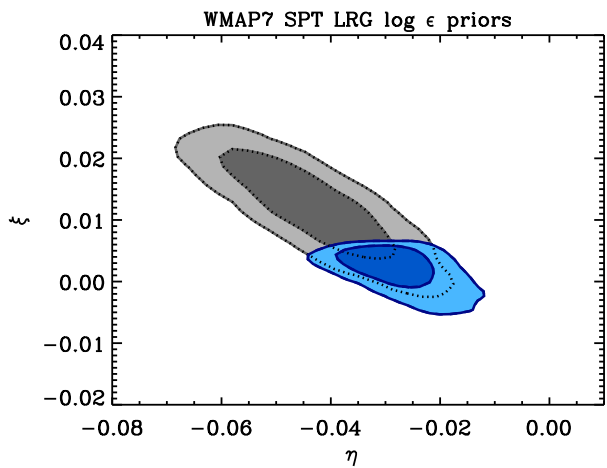}
\caption{Constraints on $\eta_\ast$ and $\xi_\ast$. These constraints were obtained by using data from WMAP7, SPT and SDSS LRGs after marginalizing over all other parameters with a log prior on $\epsilon_\ast$. Contours correspond to $68$ and $ 95$\% joint confidence levels. The blue contours correspond to discarding models which don't achieve $N_\mathrm{min}= 24$ $e$-folds from the moment at which the pivot scale exits the horizon, the gray contours correspond to $N_\mathrm{min}= 12$ $e$-folds.}
\label{plot:efolds}
\end{figure}

To determine the impact of $N_\mathrm{min}$ on our results we also performed  runs with $N_\mathrm{min}= 24$.   Figure \ref{plot:efolds} shows the constraints on $\eta_\ast$ and $\xi_\ast$ for  $N_\mathrm{min}= 12$ (grey contours) and $N_\mathrm{min}= 24$ (blue contours). As expected,  requiring fewer $e$-folds typically allows for a broader range in the slow roll parameters, especially for $\xi_\ast$ \cite{Peiris:2006sj}. Note also that the region allowed with  $N_\mathrm{min}= 12$ but excluded with $N_\mathrm{min}=24$ has a reasonably high likelihood. Consequently, excluding this region lowers the computed evidence for the ``$\epsilon_\ast$, $\eta_\ast$, $\xi_\ast$'' model, further  disfavoring the presence of $\xi_\ast$ in the parameter set. Figure \ref{plot:efolds} also shows that the posterior distribution for $\xi_\ast$ is truncated by the  $e$-folds prior with $N_\mathrm{min}= 24$, while it more closely resembles an ellipse with $N_\mathrm{min}=12$.
For our $N_\mathrm{min}\approx 24$ runs, the log of the evidence ratios increased by $\sim 1$, since in this case the $N_\mathrm{min}$ prior truncates a high-likelihood region with somewhat large values of $\xi_\ast$. 

\subsection{The Profile Likelihood Ratio.}

Given an $n$-parameter model, the $n$-dimensional likelihood encodes the information content of the data.  Often, in parameter estimation problems, we are primarily interested in the confidence intervals for a single parameter.  Using Bayes theorem, we can promote the likelihood to a probability density function (the posterior)  by multiplying by the prior probability density, and then integrating  -- or ``marginalizing'' the posterior over the remaining $n-1$  parameters. Confidence intervals obtained from the marginalized posterior  thus depend on the prior. 

Comparing the ``$\epsilon_\ast$ only'', ``$\eta_\ast$ only'' and ``$\epsilon_\ast$, $\eta_\ast$'' models in Section \ref{sec:evidence} we see that these results are strongly prior dependent. However, these priors are phenomenologically-motivated, insofar as they are not derived from fundamental physical considerations, so unlike a set of physically-derived model priors, we cannot be confident that these priors are appropriate. Thus, at least with current data, the answers to the model selection problems for this specific subset of models cannot be considered definitive.  

Consequently, it is worth investigating statistics which rely only on the likelihood and are thus prior-independent, even if these do not provide a fully consistent model selection criterion, and we now discuss the Profile Likelihood Ratio (PLR) (see e.g. Refs. \cite{PLRWilks,Reid:2009nq,GonzalezMorales:2011ty}) and turn to the Akaike Information Criterion (AIC) \cite{Akaike74} in the following subsection. The PLR  is obtained by taking the maximum value of the likelihood for fixed values of the interesting parameter;  it is the ratio of the conditional to the unconditional maximum likelihood.   This  is a straightforward  generalization of the delta chi-square ($\Delta \chi^2$) for a multidimensional likelihood in the case that we only need constraints on a single parameter: under certain regularity conditions the distribution of $-2 \ln$ PLR converges to a chi-square distribution \cite{PLRWilks}. 

By construction, the PLR statistic is prior-independent and has an interpretation similar to that of the $\Delta \chi^2$ where  an effective chi-square is identified with  $-2 \ln {\cal L}\,$.  
This analogy is exact when the likelihood is Gaussian. However, one must keep in mind that the confidence intervals may not have strict frequentist coverage, especially if the likelihood is far from Gaussian.
With this caveat in mind, best fit values and their confidence intervals can be obtained from the  $\Delta \chi^2$, or the PLR.  If the best fit parameter differs from zero at $>n$--$\sigma$ one reports an $n$--$\sigma$ ``evidence'' for that quantity.  This ``evidence'' should not be confused with the Bayesian evidence (also called evidence for short).  Being solely a description of the likelihood, the PLR is not a self-consistent model-comparison statistic, but allows us to investigate whether preferences for extra parameters  in posterior confidence intervals are driven by the data or by the prior.
  
  The use of the PLR in cosmology  is relatively recent \cite{Reid:2009nq,GonzalezMorales:2011ty}. It can be simply computed from a \MultiNest\ output using the following algorithm. 
Assume we have $m$ uninteresting parameters (which are marginalized over in the standard Bayesian approach)  and one parameter, $\beta$, on which we want to report constraints. For each value of $\beta$, we find the maximum likelihood value ${\cal L}_{\beta}$, for all values of the other $m$ parameters. We then compute $\ln ({\cal L}_{\beta}/{\cal L}_{\rm max})$, where ${\cal L}_{\rm max}$ is the overall maximum likelihood, which we identify with the maximum likelihood found by \MultiNest\ during its exploration of the full-dimensional prior. We then use  the pseudo-chisquare defined as $\ln ({\cal L}_{\beta}/ {\cal L}_{\rm max}) = 1/2 \chi^2$, so that $\Delta \ln({\cal L}_{\beta} /{\cal L}_{\rm max}) = 0.5$ and  2 correspond to the 68.3\% and 95.4\% confidence regions respectively. This quantity is, at least in principle, independent of priors. 

The PLR is useful in testing nested models: cases where a more complex model has one extra parameter ($\beta$), compared to a simpler model in which $\beta$ is fixed. For instance, in the three-parameter model we can identify  $\beta$ with $\xi_\ast$; in the simpler model $\xi_\ast \equiv 0$.  

Since \MultiNest\ uses the uncertainty on the Bayesian Evidence integral as a criterion for convergence, the computed PLR may be noisy.  The \MultiNest\ runs terminate when the evidence integral converges -- and this convergence  criterion  does not guarantee that the PLR  has also converged.  This can be a problem in practice, and the resulting numerical noise in the computed PLR is visible in some of the plots we present. 

\begin{figure}[ht]
\hspace{-4em}\includegraphics[width=0.5\textwidth]{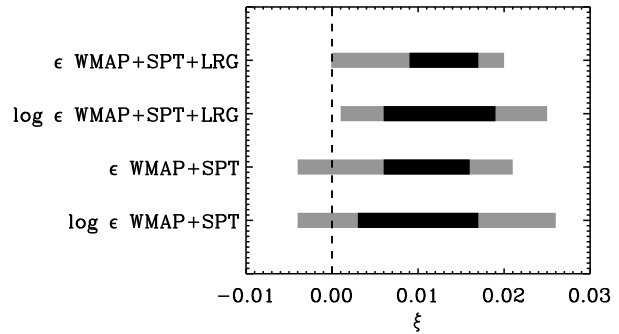}
\caption{Profile likelihood ratio-derived constraints on  the $\xi$ parameter.  Black lines show the 1-$\sigma$ intervals and grey lines the 2-$\sigma$ intervals. The dashed vertical line is a reference indicating the zero point. This result is consistent with Fig. \ref{plot:Constraints}  and indicates that the likelihood favors non-zero $\xi_*$ at about the 2-$\sigma$ level. }
\label{fig.PLRxi}
\end{figure}

\begin{figure}[ht]
\hspace{-4em}\includegraphics[width=0.5\textwidth]{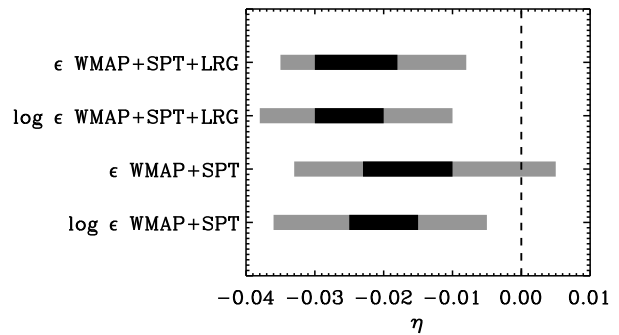}
\caption{Profile likelihood ratio-derived constraints on  the $\eta$ parameter. Black lines show the 1-$\sigma$ intervals and grey lines the 2-$\sigma$ intervals. The dashed vertical line is a reference indicating the zero point. From this plot we conclude that including $\eta_\ast$ as an extra parameter improves the likelihood such that $\eta_\ast = 0$ is excluded, at about the 2-$\sigma$ limit.}
\label{fig.PLReta}
\end{figure}

In Fig. \ref{fig.PLRxi} we show the PLR-derived constraints on  the $\xi_\ast$ parameter. From this plot we conclude that including $\xi_\ast$ as an extra parameter improves the likelihood so that $\xi_\ast = 0$ is excluded, at about the 2-$\sigma$ limit when the LRG data are included. This should be compared with Fig.~\ref{plot:xiConstraints} where the joint 2D posterior is plotted. Recall that the 2-$\sigma$ 1D marginalized constraints  are close to the projection of  2D joint 1-$\sigma$ (see e.g. Ref. \cite{numrec}). In both the (prior-dependent) posterior constraints and (prior-independent) PLR case, $\xi_\ast=0$ is just  outside the 2-$\sigma$ limit, indicating that the likelihood marginally prefers a positive $\xi_\ast$. This mild preference in the likelihood is not sufficient for the Bayesian evidence to require the inclusion of an extra parameter.

In Fig. \ref{fig.PLReta} we show the PLR-derived constraints on  the $\eta_\ast$ parameter. The analysis using a log prior on $\epsilon_\ast$ prefers a non-zero value of $\eta_\ast$, while repeating the analysis with uniform priors weakens these results. Since the PLR is supposed to be prior-independent, we interpret this difference as being due to numerical effects in the computation of the PLR, as we will discuss shortly.

First, let us compare these results with Fig.~\ref{plot:Constraints} where the  joint 2D posterior is plotted. In both cases, for the combination WMAP+SPT+LRG, $\eta_\ast=0$ is outside the 2-$\sigma$ limit, indicating that the likelihood prefers a negative $\eta_\ast$. The PLR  for $\epsilon_\ast$ (not shown) indicates that $\epsilon_\ast=0$ is always within the 1-$\sigma$ error-bar;  the PLR distribution is one-sided, decreasing as $\epsilon_\ast$ increases. Again, this is consistent with Fig.~\ref{plot:Constraints}. The fact that only upper limits on $\epsilon_\ast$ can be placed reflect the fact that there is no detection of tensor modes. But  $n_s<1$ is needed to fit the data, driving the likelihood to favor negative $\eta_\ast$ values (see Eq.~\ref{eq:ns}).
 
 The PLR results for $\xi_\ast$ (Figure \ref{fig.PLRxi}) are largely independent of the prior, as expected, but some prior dependence can be appreciated in the PLR for $\eta_\ast$ (Figure \ref{fig.PLReta}), particularly on the side of the errorbars towards zero (\emph{i.e.} large $\eta_\ast$). In fact, there is a correlation between $\epsilon_\ast$ and $\eta_\ast$, with $\eta_\ast$ becoming less negative for large $\epsilon_\ast$ (Figs. \ref{plot:Constraints} and \ref{plot:wmapConstraints}). A log prior on $\epsilon_\ast$ undersamples the likelihood for large $\epsilon_\ast$, thus underestimating the upper limit of $\eta_\ast$ in the PLR.  The effect is stronger for the cases  where larger values of $\epsilon_\ast$ are allowed by the data, where the log prior severely penalizes the sampling. This undersampling effect is much smaller for the case WMAP+SPT+LRG --which places tighter constraints on $\epsilon_\ast$-- compared to the CMB-only case.

\subsection{The Akaike information criterion}

Let us now turn to the Akaike information criterion. The Bayesian evidence is the only model-selection statistic with a self-consistent probabilitistic interpretation \cite{Cox:1946}. However, only relatively recently has it become possible to actually perform the required computationally-intensive numerical integrals in a reasonable time, in order to evaluate the Bayesian evidence in most practical applications in cosmology; suitable numerical techniques became available only recently \cite{Skilling:2004,Feroz:2007kg, Feroz:2008xx}.
Before then, approximate model-selection criteria were used. A popular example is the  Akaike information criterion (AIC) \cite{Akaike74} which has the advantage of being extremely easy and fast to compute.  
The AIC is based on the Kullback-Leibler information entropy \cite{Takeuchi:1999cu,BurnhamAnderson} and is defined as
\begin{equation}
{\rm AIC}\equiv  - 2\ln {\cal L}_{\rm max} +2k\,,
\label{eq:aic}
\end{equation}
where ${\cal L}_{\rm max}$ is the maximum likelihood achievable by the model and $k$ is the number of model parameters.  

We present the results from the AIC statistic for the models and data combinations considered here, in order to allow a direct comparison with past work where AIC was used; this makes it possible to gauge (albeit approximately) how much the new data have improved constraints on models.  A comparison of AIC with the results from Bayesian evidence ratios  can also be used to estimate how well the approximations involved in the AIC  work.  In fact Ref.~\cite{Liddle:2007fy} (which introduced the AIC to cosmology) found that, with the CMB data available at the time, the AIC and Bayesian evidence gave significantly different conclusions.

\begin{table}
\begin{tabular}{|c|c|c|}
\hline
Model & $\Delta {\rm AIC}$ & Probability ratio
\\ \hline
$\epsilon_\ast$, $\eta_\ast$, $\xi_\ast$ & $0.069$ & $0.97$
\\ \hline
$\epsilon_\ast$, $\eta_\ast$ & $2.7$ & $0.26$
\\ \hline
$\eta_\ast$ only & $0$ & $1$
\\ \hline
$\epsilon_\ast$ only & $5.9$ & $0.051$
\\ \hline
\end{tabular}
\caption{Akaike information criterion for the models considered in this work, using WMAP+SPT+LRG data and uniform priors on $\epsilon_\ast$. The first column lists the models. The second column lists the difference between the AIC for each model computed with equation \eqref{eq:aic} and the lowest AIC, which happens to be the one for the ``$\eta_\ast$ only'' model. The third column lists the probability ratio for minimizing information loss between each model and the ``$\eta_\ast$ only'' model. The numerical uncertainty on the AIC was estimated to be $\sim 0.6$.}
\label{tab:flatAIC}
\end{table}

\begin{table}
\begin{tabular}{|c|c|c|}
\hline
Model & $\Delta {\rm AIC}$ & Probability ratio
\\ \hline
$\epsilon_\ast$, $\eta_\ast$, $\xi_\ast$ & $0.75$ & $0.69$
\\ \hline
$\epsilon_\ast$, $\eta_\ast$ & $2.0$ & $0.37$
\\ \hline
$\eta_\ast$ only & $0$ & $1$
\\ \hline
$\epsilon_\ast$ only & $5.7$ & $0.057$
\\ \hline
\end{tabular}
\caption{Akaike information criterion for the models considered in this work, using WMAP+SPT+LRG data and log priors on $\epsilon_\ast$. The columns are as in Table \ref{tab:flatAIC}. The numerical uncertainty on the AIC was estimated to be $\sim 0.6$.}
\label{tab:logAIC}
\end{table}

 Tables \ref{tab:flatAIC}  and \ref{tab:logAIC}  present the AIC computed according to  Eq.~\ref{eq:aic} for each model, with a uniform and logarithmic prior on $\epsilon_\ast$ respectively. The numerical uncertainty on the AIC values listed here was estimated to be $\sim 0.6$ (uncertainty in the maximum likelihood $\sim 0.3$). Given this level of numerical uncertainty the two tables are fully consistent with each other. Note that the model with lowest AIC is the one in which we vary only over $\eta_\ast$, and it is  therefore taken as reference. The ratio $R$ of probabilities of minimizing the information loss between two models can be estimated via $R \approx \exp[({\rm AIC}_1 - {\rm AIC}_2)/2]$. We list this ratio for each model with respect to the reference  ``$\eta_\ast$ only'' model.    As expected the ``$\epsilon_\ast$, $\eta_\ast$, $\xi_\ast$'' model has the highest likelihood, but it is penalized in the AIC for having more parameters than the other models, making it virtually equivalent to the ``$\eta_\ast$ only'' case.
This is consistent with the  PLR analysis; see Figure \ref{fig.PLRxi}  where the maximum likelihood value for $\xi_\ast$ is $\sim 0.01$ but the error bars extend almost to zero. The maximum likelihoods for the ``$\epsilon_\ast$, $\eta_\ast$'' and ``$\eta_\ast$ only'' models are  close to one another (given our level of numerical uncertainty) but the ``$\epsilon_\ast$, $\eta_\ast$'' is penalized for having an additional parameter.

The AIC for the ``$\epsilon_\ast$ only'' model is considerably larger than that for the ``$\eta_\ast$ only'' model.  Both models have the same number of parameters, but the likelihood for the latter model is larger at its maximum,  implying that it is a better fit to the data. 
This is in agreement with the PLR findings: a model with $\eta_\ast=0$ is disfavored by the likelihood at  about the 2-$\sigma$ level when using the WMAP+SPT+LRG data combination (see Figure \ref{fig.PLReta}), and is again consistent with the results shown in Figures \ref{plot:Constraints} and \ref{plot:wmapConstraints}.

The AIC is sometimes used as a heuristic model-selection statistic, with a penality involving the number of model parameters going some way towards implementing Occam's razor.  We should note here that comparing it to the results from the Bayesian evidence shows that this penalty is not sufficiently conservative. By using the AIC as a model-selection statistic, we would have wrongly concluded that the ``$\epsilon_\ast$, $\eta_\ast$, $\xi_\ast$'' model is to be preferred over the ``$\epsilon_\ast$, $\eta_\ast$'' model. While the AIC leads to the correct conclusion that the single parameter ``$\eta_\ast$ only'' model is preferred over the ``$\epsilon_\ast$, $\eta_\ast$'' model, the AIC contradicts the evidence in concluding that ``$\epsilon_\ast$ only'' is disfavored with respect to the two-parameter model.

\section{Summary and conclusions}\label{sec:conclusions}

In this paper we have applied Bayesian model selection to slow roll reconstruction of the inflationary dynamics, as described by equation \eqref{eq:SRRdef} truncated at different orders. We investigated whether current cosmological data require the inclusion of high-order slow roll parameters (e.g. $\xi$), or whether they can be adequately described by just the lowest-order slow roll parameters (\emph{i.e.} $\epsilon$ and $\eta$). 

The self-consistent statistic to answer this question is the Bayesian evidence. In order to compute it efficiently we used the publicly available \MultiNest\ \cite{Feroz:2008xx,Feroz:2007kg}  extension to \CosmoMC\ \cite{Lewis:2002ah}. \MultiNest\ also computes posterior distributions that can be used for parameter estimation; this enables us to constrain the slow roll parameters for each of the models as shown in Figures \ref{plot:Constraints} and \ref{plot:xiConstraints}. We compared models where the slow roll expansion is truncated at different orders, as described in section \ref{sec:models}, using the latest compilation of CMB and large scale structure data. The slow roll approximation can induce systematic biases which are relevant for imminent precision datasets, such as CMB data from the Planck satellite. This problem can be avoided through the numerical solution of the mode equations, for which we used the publicly available {\ModeCode}\  \cite{Mortonson:2010er}.

Our approach was to treat the slow roll expansion as a phenomenological description of the inflationary dynamics, valid only during the period in which the scales that are accessible to observations evolve and freeze out. We made no assumptions about reheating or the evolution of the universe after the end of inflation, so the prior only excludes models unable to  produce sufficient inflation to account for the generation of perturbations on physical scales directly accessible to observations. This roughly corresponds to requiring that a given model ensure at least $N_\mathrm{min} \approx 12$ $e$-folds of inflation after the moment at which the pivot scale becomes of the size of the horizon.

 Even if we don't employ the slow roll approximation and impose only a minimal requirement on the duration of inflation, we see that the data prefer models which satisfy the slow roll conditions. We have simply required that models achieve just enough $e$-folds to generate the directly observable cosmological perturbations. However, if we increase the minimal number of $e$-folds  in the prior, the slow roll-allowed parameters are driven to smaller values: a longer period of inflation requires that the potential is flat over a wider field range of field values, as discussed in Ref. \cite{Easther:2006tv}. 

Slow Roll Reconstruction as implemented here will not recover sharp features like steps, or a modulated potential.  However, even if the power spectrum was generated by one of these potentials, they can only be well-described within the Hamilton-Jacobi hierarchy if a large number of higher order terms are retained, and it is not clear that the expansion would converge in this limit.  Moreover, given that the calculated evidence values demonstrate that a one-parameter model is a good fit to the data, it is not clear that models with sharp features would be favored over the simplest slow roll models, given the overall formalism of Slow Roll Reconstruction. 

The Bayesian evidence ratios are presented in Figures \ref{plot:Evidence}, \ref{plot:epsiEvidence} and \ref{plot:etaEvidence}. This analysis showed that the CMB data are well-described without the inclusion of $\xi_\ast$, and that there is  ``strong'' evidence (on the Jeffreys scale)  that the inclusion of this slow roll parameter is disfavored (\emph{i.e.}, a model with only $\epsilon_\ast$ and $\eta_\ast$  has strong evidence  compared to a model with $\epsilon_\ast$, $\eta_\ast$ and $\xi_\ast$). The inclusion of LSS LRG power spectrum slightly weakens the results but the evidence remains ``strong''. This might be due to the combination of heterogeneous data sets; for example, something as simple as a systematic mismatch in normalization  can be spuriously fit by an extra parameter. We used both logarithmic and uniform priors on $\epsilon_\ast$ and found that the numerical value for the evidence ratio depends  slightly on the form of the prior for $\epsilon_\ast$. Despite this dependence, the evidence against the inclusion of $\xi_\ast$ remains ``strong'',  making this conclusion robust to prior choice. These results are consistent with results both from minimally-parametric as well as Bayesian approaches to the reconstruction of the primordial power spectrum (for example, see Refs. \cite{Peiris:2006sj,Bridges:2008ta}).

We also compared the single-parameter models, in which only  $\epsilon_\ast$ or $\eta_\ast$ are free parameters, with models containing both parameters, finding that there is no strong indication that  more than one slow roll parameter is needed to fit the data.  
With a log $\epsilon_\ast$ prior the parameter space is heavily weighted towards small $\epsilon_\ast$, and approaches the ``Low-$\epsilon$" scenario: $\epsilon_\ast$ decouples from the hierarchy, effectively making it a nuisance parameter in the absence of a tensor detection. 
With a uniform $\epsilon_\ast$ prior, the evidence ``decisively" prefers a single parameter model over a two-parameter model, but does not indicate whether the parameter should be $\epsilon_\ast$ or $\eta_\ast$.  A detection of  (or strong upper limit on) primordial tensors would make it possible to differentiate  the ``$\epsilon_\ast$ only'', ``$\eta_\ast$ only'' and ``$\epsilon_\ast, \eta_\ast$'' models; such a constraint would also be needed in order to obtain consistent model-selection results between logarithmic and uniform priors on $\epsilon_\ast$. 
 
 Since our Bayesian results are derived from phenomenological (rather than physical) priors, we also considered a prior-independent statistic, the profile likelihood ratio (PLR), which can used to derive prior-independent confidence levels which can then be compared with the prior-dependent posterior confidence intervals. Using the PLR, we found that  the inclusion of $\xi_\ast$ as a parameter improves the likelihood just at the 2-$\sigma$ level. The inclusion of $\eta_\ast$ also improves the likelihood, so $\eta_\ast = 0$ is disfavored at slightly more than the 2-$\sigma$ level.  The PLR results are consistent with the Bayesian posterior constraints, indicating that for parameter  estimation the prior dependence is mild. 
 
In our analysis, we have uncovered a numerical pitfall in computing PLR from Bayesian samplers (see e.g. Ref. \cite{Feroz:2011bj}). Indeed, \MultiNest\ uses the uncertainty on the Bayesian Evidence as a criterion for convergence. Therefore the computed  PLR may not have the required coverage properties (\emph{i.e.} errors are underestimated) due to the fact that it may undersample some regions of parameter space which give only a small contribution to the evidence. This happens for example when using a logarithmic prior on $\epsilon_\ast$, where the large $\epsilon_\ast$ region is poorly sampled. We found that this undersampling effect can be greatly reduced by running a Markov Chain Monte Carlo analysis with much more stringent convergence criteria than the \CosmoMC\ defaults.

Finally, we computed the Akaike Information Criterion for the
different models, finding that the results from AIC were consistent
with both the PLR and the Bayesian posterior intervals.
However, unlike the PLR, the AIC is not based purely on the likelihood, and attempts to correct for model complexity by penalizing the likelihood by the number of extra parameters. Previously the AIC has been used as an approximate model-selection statistic, as it is  much easier to compute numerically than the Bayesian evidence. Comparing the model selection conclusions from the AIC with the Bayesian evidence, we see that the AIC tends to have a lower threshold for favoring the introduction of extra parameters. Thus, it tends to be less conservative than the Bayesian evidence in deciding whether the improvement in likelihood is sufficient to prefer a more complex model over a simpler one.

We expect that the SRR as implemented here will lead to significant improvements \cite{Huang:2012} in our knowledge of the inflationary dynamics when applied to future data, such as CMB data from the Planck satellite \cite{:2006uk} and the LSS power spectrum as probed by the Baryon Oscillation Spectroscopic Survey (BOSS) \cite{Schlegel:2009hj} and the Euclid satellite \cite{euclid}.

\appendix

\section{Bounds on the coefficients of the slow roll expansion.}\label{sec:bounds}

\begin{table*}[ht!]
\begin{tabular}{|c|c|c|c|c|c|c|c|}
\hline
Run & Parameter & \multicolumn{2}{|c|}{WMAP} & \multicolumn{2}{|c|}{WMAP+SPT} & \multicolumn{2}{|c|}{WMAP+SPT+LRG}
\\ \cline{3-8}
&& mean & $\sigma$ & mean & $\sigma$ & mean & $\sigma$ \\ \hline \hline
$\epsilon_\ast$ only & $\epsilon_\ast$ & - & - & $7.25\times 10^{-3}$ & $2.07\times 10^{-3}$ & $8.52\times 10^{-3}$ & $1.84\times 10^{-3}$
\\ \hline
$\log_{10} \epsilon_\ast$ only & $\log_{10} \epsilon_\ast$ & -& - & $-2.33$ & $7.94\times 10^{-1}$ & $-2.11$ & $1.98\times 10^{-1}$
\\ \hline
$\eta_\ast$ only & $\eta_\ast$ & - & - & $-2.23\times 10^{-2}$ & $5.40\times 10^{-3}$ & $2.55\times 10^{-2}$ & $4.85\times 10^{-3}$
\\ \hline
$\epsilon_\ast$, $\eta_\ast$ & $\epsilon_\ast$ & $6.71\times 10^{-3}$ & $4.96\times 10^{-3}$ & $4.20\times 10^{-3}$ & $3.39\times 10^{-3}$ & $3.21\times 10^{-3}$ & $2.76\times 10^{-3}$
\\ \cline{2-8}
& $\eta_\ast$ & $3.76\times 10^{-3}$ & $1.89\times 10^{-2}$ & $-1.10\times 10^{-2}$ & $1.08\times 10^{-2}$ & $-1.72\times 10^{-2}$ & $8.42\times 10^{-3}$
\\ \hline
$\log_{10} \epsilon_\ast$, $\eta_\ast$ & $\log_{10} \epsilon_\ast$ & $-6.19$ & $2.22$ & $-6.30$ & $2.16$ & $-6.30$ & $2.12$
\\ \cline{2-8}
& $\eta_\ast$ & $-1.67\times 10^{-2}$ & $7.79\times 10^{-3}$ & $-2.15\times 10^{-2}$ & $5.75\times 10^{-3}$ & $-2.49\times 10^{-2}$ & $4.93\times 10^{-3}$
\\ \hline
$\epsilon_\ast$, $\eta_\ast$, $\xi_\ast$ & $\epsilon_\ast$ & $6.20\times 10^{-3}$ & $5.02\times 10^{-3}$ & $4.40\times 10^{-3}$ & $3.77\times 10^{-3}$ & $3.13\times 10^{-3}$ & $2.75\times 10^{-3}$
\\ \cline{2-8}
& $\eta_\ast$ & $-6.13\times 10^{-3}$ & $2.52\times 10^{-2}$ & $-2.27\times 10^{-2}$ & $1.50\times 10^{-2}$ & $-3.06\times 10^{-2}$ & $1.13\times 10^{-2}$
\\ \cline{2-8}
&$\xi_\ast$ & $4.15\times 10^{-3}$ & $7.43\times 10^{-3}$ & $7.63\times 10^{-3}$ & $4.58\times 10^{-3}$ & $8.86\times 10^{-3}$ & $4.09\times 10^{-3}$
\\ \hline
$\log_{10} \epsilon_\ast$, $\eta_\ast$,  & $\log_{10} \epsilon_\ast$ & $-6.58$ & $2.11$ & $-6.38$ & $2.10$ & $-6.44$ & $2.12$
\\ \cline{2-8}
$\xi_\ast$ & $\eta_\ast$ & $-3.86\times 10^{-2}$ & $2.03\times 10^{-2}$ & $-3.90\times 10^{-2}$ & $1.16\times 10^{-2}$ & $-4.33\times 10^{-2}$ & $1.00\times 10^{-2}$
\\ \cline{2-8}
& $\xi_\ast$ & $1.04\times 10^{-2}$ & $9.07\times 10^{-3}$ & $1.04\times 10^{-2}$ & $6.19\times 10^{-3}$ & $1.21\times 10^{-2}$ & $5.64\times 10^{-3}$
\\ \hline
\end{tabular}
\caption{Mean values and 68\% CL obtained for the relevant parameters of each model, after marginalizing over all other parameters. The designations of the models are as listed in section \ref{sec:models}; we also specify whether a log or a uniform prior was assumed on $\epsilon_\ast$.}
\label{tab:parameters}
\end{table*}
In Table \ref{tab:parameters} we list the mean values and 68\% confidence levels for the slow roll parameters at the pivot scale, for all the scenarios we considered. Note that the biggest improvement attained by including the SPT and LRG data is in the measurement of $\eta_\ast$, for which the data indicates a non-zero value, but this improvement is not highly significant ($\sim 1\sigma$).  Note also that when $\xi_\ast$ is included, it is measured to be very small, with large uncertainties; this already hints at the fact that its inclusion is not required in order to fit the data.

\section*{Acknowledgements}
LV, JN and CW are supported by FP7-IDEAS-Phys.LSS 240117; LV  and CW  are also supported by  MICINN grant AYA2008-03531. HVP is supported in part by Marie Curie grant MIRG-CT-2007-203314 from the European Commission, and by STFC and the Leverhulme Trust.  RE is partially supported by the United States Department of Energy (DE-FG02-92ER-40704) and  National Science Foundation (CAREER-PHY-0747868).  This work began during a course   at the ICC in Barcelona, supported by the ERC  grant  FP7-IDEAS-Phys.LSS 240117.  We acknowledge the use of the Legacy Archive for Microwave Background Data (LAMBDA). Support for LAMBDA is provided by the NASA Office of Space Science.


\end{document}